\documentclass[12pt]{iopart}
\usepackage{graphics,iopams, color}

\makeatletter 
\def\@fnsymbol#1{^{\thefootnote}\relax}
\makeatother

\begin{document}

\title{Orbital Ordering and Magnetic Interactions in BiMnO$_3$}

\author{I V Solovyev$^1$, Z V Pchelkina$^2$}

\address{$^1$ Computational Materials Science Center,
National Institute for Materials Science, 1-2-1 Sengen, Tsukuba,
Ibaraki 305-0047, Japan}
\ead{solovyev.igor@nims.go.jp}
\address{$^2$ Institute of Metal Physics, Russian Academy of Sciences -- Ural Division,
620041 Ekaterinburg GSP-170, Russia}

\begin{abstract}
This work is devoted to the analysis of orbital patterns and related to
them interatomic magnetic interactions in
centrosymmetric monoclinic structures of BiMnO$_3$, which have been
recently determined experimentally.
First, we set up
an effective lattice fermion model for the manganese $3d$ bands and derive
parameters of this model entirely from
first-principles electronic structure calculations.
Then, we solve this model in terms of the mean-field Hartree-Fock
method and derive parameters of interatomic magnetic interactions
between Mn ions.
We argue that
although
nearest-neighbor interactions
favors the
ferromagnetism,
they compete with longer-range antiferromagnetic (AFM)
interactions, the existence of which is directly related with the peculiar
geometry of the
orbital ordering pattern realized in BiMnO$_3$ below 474 K.
These AFM interactions favor an AFM phase,
which breaks the inversion symmetry. The formation of the
AFM phase is
assisted by the orbital degrees of freedom, which tend to adjust
the nearest-neighbor magnetic interactions in
the direction, which
further stabilizes this phase.
We propose that the multiferroelectric behavior, observed in BiMnO$_3$, may be
related with the emergence of the AFM phase under certain conditions.
\end{abstract}

\pacs{75.30.-m, 77.80.-e, 75.47.Lx, 75.10.Lp}
%\maketitle

\section{\label{sec:Introduction} Introduction}

  The multiferroic compounds have recently
drawn an enormous attention
due to
promising practical applications
as well as the fundamental interest \cite{Fiebig,Khomskii,Eerenstein,Cheong}.
Such systems,
where magnetism coexists with the ferroelectricity,
could be potentially used in the
new devices aiming to transform the information in the form of the magnetization
into the electric voltage and back.
The primary goal of theorists
is to unveil the microscopic
mechanism leading to the coupling
between magnetic and electric degrees of freedom.

  The bismuth manganite (BiMnO$_3$), having highly distorted perovskite structure,
has been regarded as one of the prominent multiferroic materials.
Indeed, the ferromagnetism of BiMnO$_3$ is well established today.
The Curie temperature ($T_C$) is about 100 K.
The largest reported saturation magnetization is $3.92~\mu_B$ per one
formula unit \cite{belik_07},\footnote{
measured at the temperature of $5$ K and in the magnetic field of $5$ T.
}
which is close to $4~\mu_B$
expected for the fully saturated ferromagnetic (FM) state. Nevertheless, the saturation
magnetization
decreases rapidly with the doping in Bi$_{1-x}$Sr$_x$MnO$_3$ \cite{Chiba},
that may indicate at the proximity of yet another and apparently
antiferromagnetic (AFM) phase.

  However, the situation around the ferroelectric properties of
BiMnO$_3$ is more controversial.
There are several facts, which do support the idea that BiMnO$_3$ is not only
ferromagnetic, but also a ferroelectric material.
\begin{enumerate}
\item
The existence of ferroelectricity has been advocated by first-principles
electronic structure calculations \cite{HillRabe}, and attributed to the chemical
activity of the Bi($6s^2$) lone pairs \cite{SeshadriHill}, in an analogy with
other ferroelectric materials, such as PbTiO$_3$.
\item
According to early experimental data from electron and neutron powder diffraction,
BiMnO$_3$ was considered to have noncentrosymmetric $C2$ space group
in the entire monoclinic region \cite{atou_99, santos_02},
which is consistent with the ferroelectric behavior.
Namely,
BiMnO$_3$ undergoes two phase transitions at the temperatures of 474 and 770
K \cite{atou_99, santos_02}. The first one at 474 K takes place without
changing the monoclinic symmetry \cite{montanari_05}. The phase
transition at 770 K is monoclinic to orthorhombic \cite{montanari_05}
and was believed likely to be ferroelectric-paraelectric.
Nevertheless, it is also worth to note that
this point of view is rather controversial and according to \cite{SantosSSC}
the onset of the ferroelectric behavior is expected only around 450 K, which
in \cite{SantosSSC}
was
the point of isostructural
(i.e., monoclinic to monoclinic) phase transition.
\item
The ferroelectric hysteresis loop has been also reported in polycrystalline and
thin film samples of BiMnO$_3$ \cite{SantosSSC}, although the measured ferroelectric
polarization was small (about $0.043$ $\mu$C/cm$^2$ at 200 K).
The first principle calculations performed for
the experimental
noncentrosymmetric structure result in much higher polarization
(about 0.52 $\mu$C/cm$^2$) \cite{Shishidou_04}.
\item
Kimura {\it et al.} \cite{Kimura} observed the changes of the dielectric constant induced by the
magnetic ordering as well as by the external magnetic field near
$T_C \sim 100$ K, and attributed them to the multiferroic behavior
of BiMnO$_3$.
\item
Sharan {\it et al.} \cite{Sharan} observed the
electric-filed-induced permanent changes in the second harmonic response
from the BiMnO$_3$ thin film, and argued that these changes are consistent
with the possible presence of ferroelectricity.
\end{enumerate}

  However, there is also a growing evidence against the intrinsic ferroelectric
behavior of BiMnO$_3$.
\begin{enumerate}
\item
Recently
the crystal structure of BiMnO$_3$ was reexamined by Belik {\it et al.} \cite{belik_07}.
After careful analysis, they concluded that
both monoclinic phases observed in
BiMnO$_3$ below 770 K have \textit{centrosymmetric} space group $C2$/$c$.
If so, BiMnO$_3$ should be an \textit{antiferroelectric}, rather than
the ferroelectric material. This funding was further confirmed in the
neutron powder diffraction experiments by Montanari {\it et al.} \cite{montanari_07} who also
concluded that the crystal structure of BiMnO$_3$ is better described by the $C2$/$c$ group
in the wide range of temperatures
(10 $\leq T \leq$ 295 K) and magnetic fields (0 $\leq H \leq$ 10 T).
It is also important to note
that there are many objective difficulties
in the determination of the crystal structure of BiMnO$_3$, which are
mainly related with the strong
effect of nonstoichiometry \cite{Yokosawa}.\footnote{
In fact, the experimental situation is
complicated by the samples differences (thin films or bulk),
nonstoichiometry, effect of substrate (for the thin films), etc.
Some difficulties
and artifacts arising in the experiment for multiferroic and
magnetoelectric thin films have been discussed in \cite{EerensteinPML}.}
\item
For the related compound BiScO$_3$, both neutron powder and electron diffraction analysis
result in the centrosymmetric $C2$/$c$ space group \cite{belik_bisco3}.
\item
The structure optimization
performed by using modern methods of
electronic structure calculations revealed that
the noncentrosymmetric $C2$ structure, which has been reported earlier \cite{atou_99,santos_02},
inevitably converges to the new total energy minimum corresponding to the
$C2$/$c$ structure with zero net polarization \cite{Shishidou, spaldin_07}.
\end{enumerate}

  The goal of this work is to study of the orbital ordering
and corresponding to it interatomic magnetic interactions in the
centrosymmetric structure of BiMnO$_3$. For these purposes we construct an
effective lattice fermion model and derive
parameters of this model from first-principles
electronic structure calculations. After solution of this
model we calculate the interatomic magnetic interactions. We argue that the peculiar
orbital ordering
realized below 474 K gives rise to FM interactions
between nearest-neighbor spins which always compete with longer-range
AFM interactions. We propose that
the ferroelectric behavior
of BiMnO$_3$ can be related with the emergence of an AFM phase,
which is stabilized by these longer-range interactions
and breaks the inversion symmetry.

  Thus, according to our point of view, the multiferroic behavior of
BiMnO$_3$ is
feasible and
can be
related with competition of two
magnetic phases coexisting in a narrow energy range. The centrosymmetric FM ground state
itself
is antiferroelectric.
Nevertheless, the ferroelectricity can be observed in the
noncentrosymmetric
AFM phase, which can apparently exist under certain conditions.
Since the ferromagnetic (antiferroelectric) and antiferromagnetic (ferroelectric)
phases can be stabilized by applying the magnetic and electric field, respectively,
the magnetic moment can be switched off by the electric field and vice versa.
This constitutes our idea of multiferroic behavior of BiMnO$_3$.
We rationalize several experimental facts on the basis of this picture.

  The paper is organized as follows. In the next section
we discuss details of the
centrosymmetric crystal structure of BiMnO$_3$.
Section \ref{sec:estruc} briefly describes
results of
first principle electronic
structure calculations in the local-density approximation (LDA). The
construction of the model Hamiltonian is
addressed in Section \ref{sec:ModelParameters}. The
solution of the model Hamiltonian and
physical meaning of interatomic magnetic
interactions is discussed in Section \ref{sec:ModelAnalysis}.
The results of calculations are discussed
in Section \ref{sec:Results}.
Finally, in Section \ref{sec:Summary} we will summarize our work
and discuss how our results are related with
the observed experimental behavior of BiMnO$_3$.

\section{\label{sec:CrystalStructure} Crystal Structure and Symmetry Considerations}

\begin{figure}[h!]
\begin{center}
\resizebox{5cm}{!}{\includegraphics{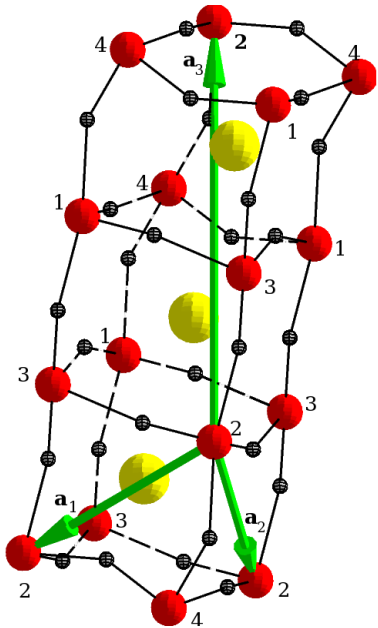}}
\end{center}
\caption{\label{fig.structure} Fragment of crystal structure of BiMnO$_3$,
presented in the form of highly distorted perovskite lattice.
The Bi atoms are indicated by the big yellow (light grey) spheres,
the Mn atoms are indicated by the medium red (dark grey) spheres, and
the oxygen atoms are indicated by the small hatched spheres.
The primitive cell includes four Mn atoms, which are indicated by the numbers.
The primitive translations are shown by arrows.
}
\end{figure}

  BiMnO$_3$ has a highly distorted perovskite structure
(figure \ref{fig.structure}).

  In our calculations we used the experimental crystal structure
for $T$$=$ 4 and 550 K
obtained by Belik {\it et al.}
The experimental structure parameters for $T$$=$ 550 K can be found in \cite{belik_07},
while the ones for $T$$=$ 4 K are unpublished data \cite{belik_pc}.\footnote{
The structure parameters for $T$$=$ 4 K are pretty close to the ones for
reported in \cite{belik_07} for
$T$$=$ 300 K.
If fact, all calculations have been performed
using the experimental crystal structure
both for
$T$$=$ 4 K and $T$$=$ 300 K.
Since both structures produce similar results,
in the following we consider only the
case of $T$$=$ 4 K.
}
The primitive translations in the original monoclinic coordinate frame
are give by
\begin{eqnarray*}
{\bf a}_1 & = & (            4.4605,           -2.8019,           -1.6748) \nonumber\\
{\bf a}_2 & = & (            4.4605, \phantom{-}2.8019,           -1.6748) \nonumber\\
{\bf a}_3 & = & (  \phantom{4.465}0, \phantom{-2.819}0, \phantom{-}9.8481)
\end{eqnarray*}
(in \AA, for $T$$=$ 4 K).

   The space group $C2/c$ has four symmetry operations:
\begin{eqnarray}
\hat{S}_1 & = & \{ E|{\bf 0} \}       \nonumber\\
\hat{S}_2 & = & \{ I|{\bf 0} \}       \nonumber\\
\hat{S}_3 & = & \{ m_y|{\bf a}_3/2 \} \nonumber\\
\hat{S}_4 & = & \{ C^2_y|{\bf a}_3/2 \},
\label{eqn:symmetry}
\end{eqnarray}
where in the notation $\{ {\cal O} | {\bf t} \}$,
${\cal O}$$=$ $E$, $I$, $m_y$, or $C^2_y$ denotes the
local symmetry operation, which is combined with the translation
${\bf t}$$=$ ${\bf 0}$ or ${\bf a}_3/2$.
Other notations are the following:
$E$ is the unity operation, $I$ is the inversion, $m_y$ is the mirror
reflection of the axis $y$, and $C^2_y$ is the $180^\circ$ rotation around $y$.

  The primitive cell of BiMnO$_3$ has four formula units.
Four Mn atoms are located at
\begin{eqnarray*}
{\rm Mn1} & : & (\phantom{2.233}0,           -1.2069, \phantom{-}2.4620) \nonumber\\
{\rm Mn2} & : & (\phantom{2.233}0, \phantom{-}1.2069,           -2.4620) \nonumber\\
{\rm Mn3} & : & (          2.2303,           -1.4009,           -0.8374) \nonumber\\
{\rm Mn4} & : & (          2.2303, \phantom{-}1.4009, \phantom{-}4.0866)
\end{eqnarray*}
(in \AA).
They can be divided in two groups (Mn1, Mn2) and (Mn3, Mn4), so that in each
group the atoms can be transformed to each other by the symmetry operations (\ref{eqn:symmetry}).
The corresponding transformation law is given in table \ref{tab:TransformationLaw}.
\begin{table}[h!]
\caption{The transformation law of four Mn atoms in BiMnO$_3$
under symmetry operations of the $C2/c$ group. Four Mn atoms are listed in the
first column. Next four columns show their images
after applying the symmetry operations
of the $C2/c$ space group.}
\label{tab:TransformationLaw}
\begin{indented}
\item[]\begin{tabular}{@{}c|cccc}
\br
     atom   & $\{ E|{\bf 0} \}$ & $\{ I|{\bf 0} \}$ & $\{ m_y|{\bf a}_3/2 \}$ & $\{ C^2_y|{\bf a}_3/2 \}$ \\
\mr
    Mn1  &    Mn1       &    Mn2           &      Mn2             &      Mn1               \\
    Mn2  &    Mn2       &    Mn1           &      Mn1             &      Mn2               \\
    Mn3  &    Mn3       &    Mn3           &      Mn4             &      Mn4               \\
    Mn4  &    Mn4       &    Mn4           &      Mn3             &      Mn3               \\
\br
\end{tabular}
\end{indented}
\end{table}

  The ferromagnetic configuration of BiMnO$_3$ has the full
$C2/c$ symmetry, which excludes any ferroelectricity.
Possible antiferromagnetic configurations can be obtained by combining the symmetry
operations (\ref{eqn:symmetry}) with the time-inversion $\hat{T}$, which flips
directions of the magnetic moments. Then, one can expect the following possibilities:
\begin{enumerate}
\item
The AFM configuration $\uparrow \uparrow \downarrow \downarrow$
(in these notations, four arrows
correspond to the directions of the magnetic moments at the Mn-sites
$1$, $2$, $3$, and $4$, respectively),
which transforms according to the original space group $C2/c$.
Similar to the FM case, this configuration exclude the ferroelectricity.
\item
The AFM configuration $\uparrow \downarrow \downarrow \uparrow$,
which
apart from the unity element $\{ E|{\bf 0} \}$, has only one
symmetry operation: $\hat{T} \otimes \{ m_y|{\bf a}_3/2 \}$.
This configuration does allow for the ferroelectricity,
and
the spontaneous polarization is expected to be perpendicular to the $y$-axis.
Once the symmetry is broken by the AFM order, the atomic position will
shift in order to minimize the total energy via magneto-elastic
interactions. In this case, the crystal symmetry is expected to be $P2$, which is
compatible with the magnetic symmetry of the AFM $\uparrow \downarrow \downarrow \uparrow$
phase.\footnote{
In the other words, the distribution of the \textit{electron density} in the AFM
$\uparrow \downarrow \downarrow \uparrow$ phase obeys the $P2$ group, while
the distribution of the \textit{magnetization density} obeys the magnetic group
in which $\{ m_y|{\bf a}_3/2 \}$ is combined with $\hat{T}$.
}
Thus, the ferroelectric behavior in the
$\uparrow \downarrow \downarrow \uparrow$ phase is driven by the magnetic breaking of
the inversion symmetry. A similar scenario of appearance
of the ferroelectricity has been recently considered for other manganese oxides:
HoMnO$_3$ \cite{Sergienko,Picozzi} and TbMn$_2$O$_5$ \cite{Wang}.
\end{enumerate}
Other combinations of $\hat{T}$ with the symmetry operations (\ref{eqn:symmetry})
will lead to unphysical solutions, where the local magnetic moments will vanish
in one of
the Mn-sublattices.
Although such configurations are formally allowed by the symmetry, they
clearly conflict with intraatomic Hund's first rule and are expected to have
much higher energy.\footnote{
Note that in manganites, the intraatomic exchange coupling
$J$ is about 0.9 eV and the local magnetic moment $M$ is about
$4 \mu_B$. Therefore, the Hund energy, $-$$\frac{1}{4}JM^2$, is expected to be
about $3.6$ eV per one Mn site.
}
We have also considered two ferrimagnetic configurations
$\uparrow \downarrow \downarrow \downarrow$ and
$\downarrow \downarrow \uparrow \downarrow$,
which do not have any symmetry.
In this case, the spontaneous polarization may have an arbitrary direction.

  Below 474 K, the MnO$_6$ octahedra are strongly distorted.
Around each Mn site,
there are three inequivalent pairs of Mn-O bonds.
At $T$$=$ 4 K,
the values of the Mn-O bondlengths are
( 1.899 \AA, 1.997 \AA, 2.189 \AA) and
( 1.930 \AA, 1.940 \AA, 2.230 \AA), around the sites Mn1 and Mn3, respectively.
In the first approximation, one can say that there are four short
Mn-O bonds and two long Mn-O bonds. This distortion leads to the
preferential population of the $e_g$ orbitals of the $z^2$ symmetry,
aligned along the long Mn-O bonds.
The difference between the longest and shortest Mn-O distances
around the sites Mn1 and Mn3 is 0.290 \AA~and 0.300 \AA, respectively.
This distortion is substantially
relived above 474 K. For example, at $T$$=$ 550 K, the Mn-O bondlengths
around the sites Mn1 and Mn3 are ( 2.011 \AA, 2.032 \AA, 2.112 \AA) and
( 1.913 \AA, 2.024 \AA, 2.106 \AA), respectively, and the
difference between the longest and shortest Mn-O distances
around the sites Mn1 and Mn3 is reduced till
0.101 \AA~and 0.193 \AA, respectively.
Thus, the reduction is the most dramatic around the site Mn1.

\section{Electronic Structure in the Local-Density Approximation}
\label{sec:estruc}

\begin{figure}[h!]
\begin{center}
\resizebox{7cm}{!}{\includegraphics{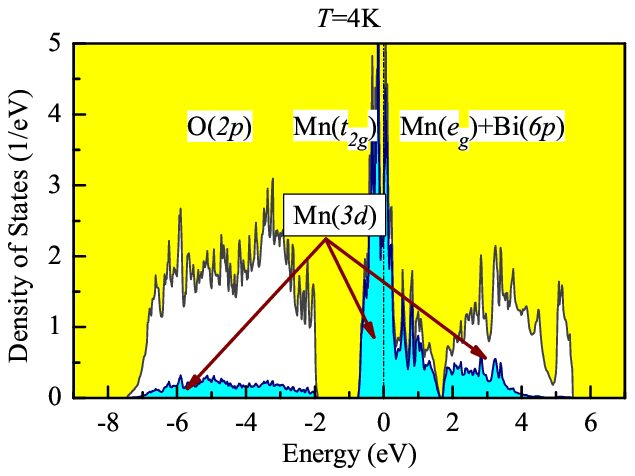}}
\resizebox{7cm}{!}{\includegraphics{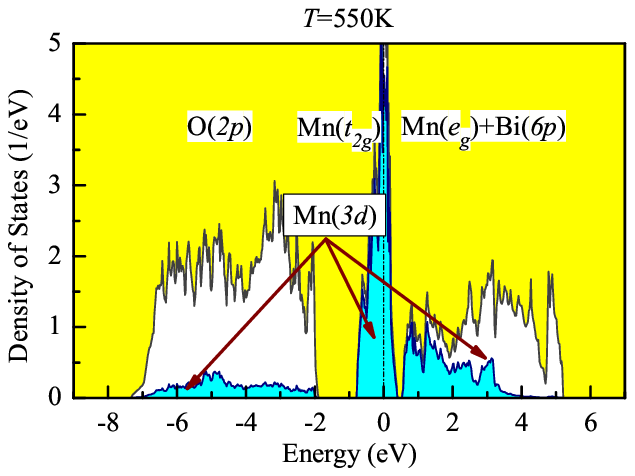}}
\end{center}
\caption{\label{fig.DOS} Total and partial densities of states
as obtained for the low-temperature (left) and high-temperature (right)
monoclinic structures of BiMnO$_3$
in the local-density approximation.
The shaded area shows contributions of the manganese $3d$ states.
Other symbols show the positions of the main bands.
The Fermi level is at zero energy.}
\end{figure}

  First, we calculate
the electronic structure corresponding to the low and high temperature
structure of BiMnO$_3$ in the local density approximation
(LDA) by using linear-muffin-tin-orbital (LMTO) method \cite{LMTO1,LMTO2,LMTO3}.
The atomic spheres radii
and some details of the LMTO basis set used in the calculation are given in table \ref{tab:LMTO}.
\Table{\label{tab:LMTO}
Details of LMTO calculations for BiMnO$_3$ at 4 and 550 K.
The notations of inequivalent oxygen atoms are the same as in \protect\cite{belik_07}.}
\br
 type of atom & LMTO basis           & \centre{2}{radii (\AA)} \\
\ns
              &                      & \crule{2}               \\
              &                      & 4 K         &     550 K \\
\mr
Bi            & 6$s$6$p$6$d$5$f$     & 1.59        & 1.54      \\
Mn1           & 4$s$4$p$3$d$         & 1.24        & 1.32      \\
Mn3           & 4$s$4$p$3$d$         & 1.26        & 1.25      \\
O1            & 2$s$2$p$             & 0.98        & 1.04      \\
O2            & 2$s$2$p$             & 0.97        & 1.02      \\
O3            & 2$s$2$p$             & 0.99        & 0.97      \\
\br
\end{tabular}
\end{indented}
\end{table}
In order to
fill the unit cell volume
and reduce the overlap between atomic spheres,
we additionally introduced 36 and 42 empty spheres
for the 4 K and 550 K structure, respectively. The resulting total
and partial densities of state are shown in figure \ref{fig.DOS}.
The oxygen band,
lying between -7 and -2 eV, is completely filled.
The electronic structure near the Fermi level is mainly
formed by the
Mn($3d$)
states.
Due to the hybridization, there is also a considerable weight of the
Mn($3d$) states in the oxygen band.
The electronic structure near the Fermi level is further split into the Mn($e_g$) and
Mn($t_{2g}$) bands by pseudocubic crystal field operating in the MnO$_6$
octahedra, although in the highly distorted monoclinic structure
there is no unique definition of the ``$t_{2g}$'' and ``$e_g$'' orbitals
since
they are always mixed by the crystal distortion.
The distortion is particularly strong in
the low-temperature phase, leading even to an overlap between
Mn($t_{2g}$) and Mn($e_g$) bands.
The Mn($e_g$) band itself is split into the low- and high-energy subbands,
lying at around 1 eV and 3 eV, respectively. There is a small gap between
these subbands at around 1.7 eV. In total, the low-energy Mn($e_g$) subband
can accommodate one electron per one formula unit of BiMnO$_3$. Therefore,
according to the formal valence argument,
in the fully polarized FM phase,
both Mn($t_{2g}$) and low-energy Mn($e_g$) bands
are expected to be
filled for the majority-spin channel, and the Fermi level
is expected to fall in the pseudogap.
The crystal distortion is somewhat released in the
high-temperature structure, that leads to the opening of a gap
between Mn($t_{2g}$) and Mn($e_g$) bands and closing the
gap between two Mn($e_g$) subbands.
The high-energy Mn($e_g$) subbands always overlap with the
Bi($6p$) band spreading from about 2 to 5 eV.
In this sense, any attempt to construct the model Hamiltonian for the
isolated Mn($3d$) bands will be an
conjugated with some additional approximations for treating the
Bi($6p$) states and their hybridization with the Mn($3d$) states.

\section{\label{sec:ModelParameters} Construction of the Model Hamiltonian}

  Our next goal is the construction of an effective
multi-orbital Hubbard-type model for the
Mn($3d$) bands, located near the Fermi level, and derivation of the parameters of this model
from the first-principles electronic structure calculations.
The method has been proposed in \cite{PRB06a}. Many details can be found
in the recent review article \cite{review2008}.

  The model itself is specified as follows:
\begin{equation}
\hat{\cal{H}}= \sum_{{\bf R}{\bf R}'} \sum_{\alpha_1 \alpha_2} t_{{\bf
R}{\bf R}'}^{\alpha_1 \alpha_2}\hat{c}^\dagger_{{\bf
R}\alpha_1}\hat{c}^{\phantom{\dagger}}_{{\bf R}'\alpha_2} + \frac{1}{2}
\sum_{\bf R}  \sum_{ \{ \alpha \} } U^{\bf R}_{\alpha_1 \alpha_2
\alpha_3 \alpha_4} \hat{c}^\dagger_{{\bf R}\alpha_1} \hat{c}^\dagger_{{\bf
R}\alpha_3} \hat{c}^{\phantom{\dagger}}_{{\bf R}\alpha_2}
\hat{c}^{\phantom{\dagger}}_{{\bf R}\alpha_4},
\label{eqn:Hmanybody}
\end{equation}
where $\hat{c}^\dagger_{{\bf R}\alpha}$ ($\hat{c}_{{\bf R}\alpha}$)
creates (annihilates) an electron in the Wannier orbital
$\tilde{W}_{\bf R}^\alpha$ centered at the Mn-site
${\bf R}$, and $\alpha$ is
a joint index, incorporating spin ($s$$=$ $\uparrow$ or $\downarrow$) and orbital
($m$$=$ $xy$, $yz$, $z^2$, $zx$, or $x^2$$-$$y^2$)
degrees of freedom.
The one-electron Hamiltonian $\hat{t}_{{\bf R}{\bf R}'}$$=
$$\| t_{{\bf R}{\bf R}'}^{\alpha_1 \alpha_2} \|$
usually includes the following contributions:
the site-diagonal part (${\bf R}$$=$${\bf R}'$) describes
the  crystal-field splitting, whereas the off-diagonal part
(${\bf R}$$\neq$${\bf R}'$) stands for transfer integrals,
describing the kinetic energy of electrons.
$$
U^{\bf R}_{\alpha_1 \alpha_2 \alpha_3 \alpha_4}
=
\int d{\bf r} \int d{\bf r}' \tilde{W}_{\bf R}^{\alpha_1 \dagger}({\bf r})
\tilde{W}_{\bf R}^{\alpha_2}({\bf r}) v_{\rm scr}({\bf r},{\bf r}')
\tilde{W}_{\bf R}^{\alpha_3 \dagger}({\bf r}') \tilde{W}_{\bf R}^{\alpha_4}({\bf r}')
$$
are
the matrix elements of \textit{screened} Coulomb interaction
$v_{\rm scr}({\bf r},{\bf r}')$, which are supposed to be diagonal with
respect to the site indices $\{ {\bf R} \}$.
The intersite matrix elements are typically
small.

  Since we do not consider here the relativistic spin-orbit interaction,
the matrix elements
$t_{{\bf R}{\bf R}'}^{\alpha_1 \alpha_2}$ are diagonal with respect to the
spin indices: i.e.,
$t_{{\bf R}{\bf R}'}^{\alpha_1 \alpha_2} = t_{{\bf R}{\bf R}'}^{m_1 m_2} \delta_{s_1 s_2}$.
The spin-dependence of
the screened Coulomb interactions
$U^{\bf R}_{\alpha_1 \alpha_2 \alpha_3 \alpha_4}$
also has the regular form:
$U^{\bf R}_{\alpha_1 \alpha_2 \alpha_3 \alpha_4} = U^{\bf R}_{m_1 m_2 m_3 m_4} \delta_{s_1 s_2} \delta_{s_3 s_4}$.
Generally, the matrix elements of
$\hat{U}^{\bf R} = \|  U^{\bf R}_{m_1 m_2 m_3 m_4} \|$
depend on the site-index ${\bf R}$.

\subsection{\label{sec:OneElectron} One-electron part}

\begin{figure}[h!]
\begin{center}
\resizebox{!}{4.5cm}{\includegraphics{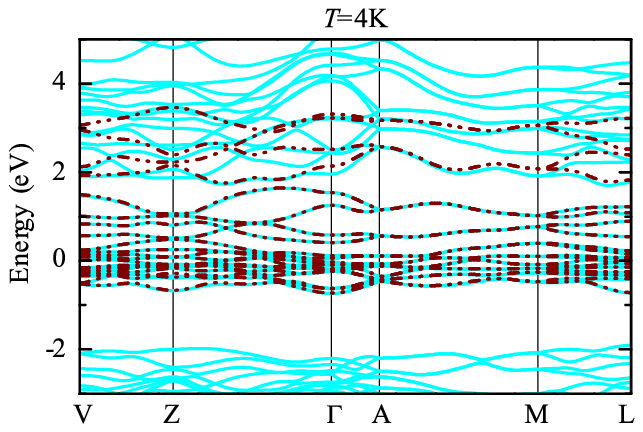}}
\resizebox{!}{4.5cm}{\includegraphics{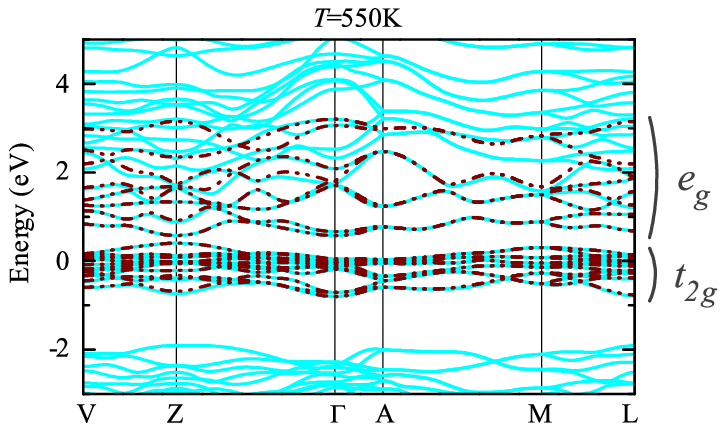}}
\end{center}
\caption{\label{fig.ek}
LDA energy bands
corresponding to the low-temperature (left) and high-temperature (right)
monoclinic structures of BiMnO$_3$ as obtained in the original
electronic structure calculations using LMTO method (solid curves) and
after the tight-binding (TB) parametrization using the downfolding method (dash-doted curves).
Notations of the high-symmetry points of the Brillouin zone are taken
from \protect\cite{BradlayCracknell}.}
\end{figure}

  The one-electron part of the model Hamiltonian (\ref{eqn:Hmanybody}) can be
constructed by using the formal downfolding method,
which is applied to the Kohn-Sham equations within LDA. The method has been
proposed in \cite{PRB06a,PRB04}. It is totally equivalent to
calculation of the matrix elements of the Kohn-Sham Hamiltonian in the
basis of Wannier functions
constructed by using the projector-operator technique \cite{PRB07}.
The advantage of the downfolding method is that by using it
one can formally bypass the construction of the Wannier functions themselves and go directly to
the calculation of the one-electron part of the model Hamiltonian.
The comparison between original LDA bands
and the ones obtained in the
downfolding method
is shown in figure \ref{fig.ek}.
Generally, the agreement is nearly perfect for the low-energy Mn($t_{2g}$) and the first four Mn($e_g$) bands.
In this region, the original electronic structure
of the LMTO method
is well reproduced after the downfolding.
However, for the upper Mn($e_g$) bands, which strongly overlap and interact with the Bi($6p$) bands,
it is virtually impossible to reproduce all details of the electronic structure
in the minimal model consisting only of the Mn($3d$) bands.\footnote{
In the other words, in order to reproduce all bands we had to expand our
Wannier basis and treat on an equal footing both Mn($3d$) and Bi($6p$) states.}
Therefore, in the upper-energy region,
the electronic structure obtained in the downfolding method is only an approximation
to the original LDA band structure.

\begin{figure}[h!]
\centering \noindent
\resizebox{12cm}{!}{\includegraphics{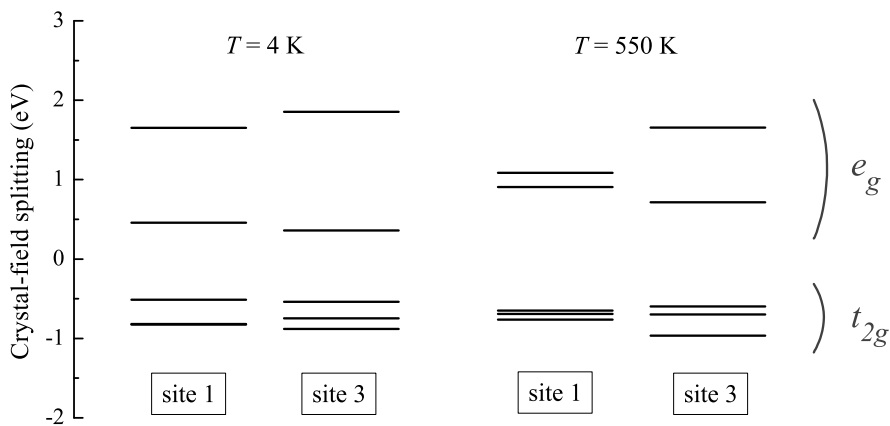}}
\caption{\label{fig.CF}
Crystal-field splitting as obtained for the low-temperature (left) and
high-temperature (right) monoclinic structures of BiMnO$_3$. Positions of the Mn sites
are explained in figure \ref{fig.structure}.}
\end{figure}

  The model parameters for the one-electron part are obtained after the
Fourier transformation of the downfolded Hamiltonian to the real space.
In this case, the site-diagonal part of $\hat{t}_{{\bf R}{\bf R}'} = \| t_{{\bf R}{\bf R}'}^{m_1 m_2} \|$
describes the crystal-field splitting.
The splitting of the $e_g$ levels in the low-temperature phase
is particularly strong, being of the order of 1.5 eV (figure \ref{fig.CF}).
As we will see below, the crystal-field
effects will lead to a peculiar type of the orbital ordering,
which will be mainly responsible for the magnetic properties of BiMnO$_3$.
This crystal-field splitting is mainly related with the change of the hybridization
(or the covalent mixing)
in different bonds of the distorted perovskite structure, which after the downfolding
gives rise to
the site-diagonal contributions in the model Hamiltonian. The nonsphericity of the Madelung potential,
which plays a crucial role in the $t_{2g}$ compounds \cite{MochizukiImada,PRB06b},
is considerably smaller
then the effects of the covalent mixing in the $e_g$ bands
and can be neglected.
In the high-temperature phase, the crystal-field splitting shrinks only in
one of the sublattices, formed by the Mn atoms `1' and `2' in figure \ref{fig.structure}.
In the second sublattice, formed by the atoms `3' and `4', the $e_g$-level splitting
remains large, being of the order 1 eV.

\begin{figure}[h!]
\begin{center}
\resizebox{7cm}{!}{\includegraphics{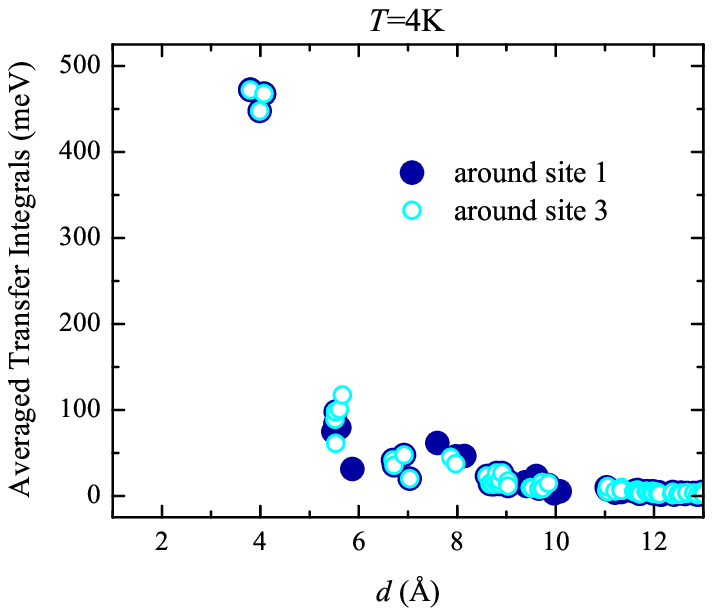}}
\resizebox{7cm}{!}{\includegraphics{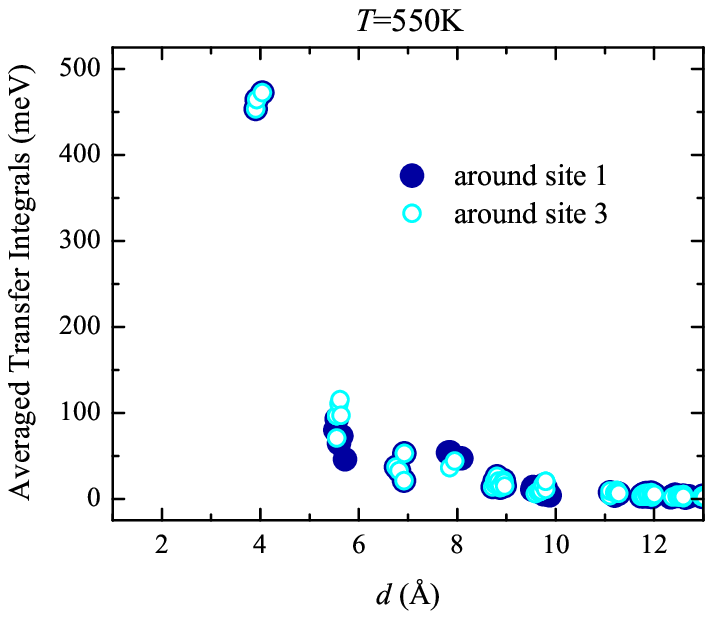}}
\end{center}
\caption{\label{fig.transfer}
Distance-dependence of averaged transfer integrals,
$\bar{t}_{{\bf RR}'}(d) = \left( \sum_{m_1 m_2} t^{m_1 m_2}_{{\bf RR}'}
t^{m_2 m_1}_{{\bf R}'{\bf R}} \right)^{1/2}$,
as obtained for the
low-temperature (left) and high-temperature (right) monoclinic
structures of BiMnO$_3$.
The values around two inequivalent
Mn sites are shown by closed and open symbols.
Positions of the Mn sites
are explained in figure \ref{fig.structure}.}
\end{figure}

   Because of complexity of the transfer integrals in the monoclinic
structure, it is rather difficult to discuss the behavior of individual
matrix elements of $\| t^{m_1 m_2}_{{\bf RR}'} \|$.
Nevertheless,
some useful information can be obtained from the analysis of
\textit{averaged} parameters
$$
\bar{t}_{{\bf RR}'}(d) = \left( \sum_{m_1 m_2} t^{m_1 m_2}_{{\bf RR}'}
t^{m_2 m_1}_{{\bf R}'{\bf R}} \right)^{1/2},
$$
where
$d$ is the distance between Mn-sites ${\bf R}$ and ${\bf R}'$.
All transfer integrals are localized and practically restricted by the nearest
neighbors at around 4\AA~(figure \ref{fig.transfer}).
The longer-range interactions are considerably smaller.

\subsection{\label{sec:screenedU} Screened Coulomb interactions}

  The matrix elements of screened Coulomb interactions in the Mn($3d$) band, $\hat{U}^{\bf R}$,
can be computed in two steps \cite{PRB06a,review2008}.
First, we perform the conventional constrained LDA calculations \cite{Gunnarsson1989}, and derive
parameters of the on-site Coulomb interaction
$u = 10$ eV and the intraatomic exchange interaction $j = 1$ eV.
These parameters are certainly too large and only weekly depend
on the crystal environment of Mn-atoms in the solid.
They include several important channel of screening: for example,
the screening
of $3d$-interactions by other electrons and
the screening caused by relaxation
of the atomic $3d$ wavefunctions are already included in the definition of $u$ and $j$.
However, this is not the whole screening and what the constrained LDA
typically cannot do is to treat the so-called self-screening caused
\textit{by the same $3d$ electrons},
which participate in the formation of other bands
due to the hybridization effects \cite{review2008}.
The major contribution comes from the O($2p$) band, which has a large weight
of the Mn($3d$) states (figure \ref{fig.DOS}).
This channel of screening can be efficiently
taken into account in the random-phase approximation (RPA)
by starting from interaction parameters obtained in the constrained LDA \cite{PRB06a}:
$$
\hat{U} = \left[1- \hat{u} \hat{P} \right]^{-1} \hat{u},
$$
where $\hat{P}$ is the static polarization matrix in RPA,
calculated in the basis of atomic $3d$ orbitals,\footnote{
In the present context, the ``atomic orbitals'' mean the muffin-tin
orbitals of the LMTO method \cite{LMTO1,LMTO2,LMTO3}.}
and $\hat{u}$ is the $5$$\times$$5$$\times$$5$$\times$$5$ matrix
of Coulomb interaction in the atomic limit.
For each transition-metal site, $\hat{u}$
can be obtained from the parameters $u$ and $j$ by using a
regular procedure, which is typically adopted in the LDA$+$$U$ method \cite{review2008}.
The polarization matrix $\hat{P}$ is computed by using the LDA band structure.
Nevertheless, in order to simulate the electronic structure close to the
saturated ferromagnetic ground state, we have used different Fermi
levels for the majority- and minority-spin channels in the process of calculation of $\hat{P}$.
Namely, for the minority-spin channel, it was assumed that the Mn($3d$) band is empty and
the Fermi level has been placed right after
the O($2p$) band (i.e., around -1.5 eV in figure \ref{fig.ek}), while
for the majority-spin channel, it was assumed that the Mn($3d$) band accommodates
all sixteen electrons (per four formula units).
Meanwhile, we switch off all contributions to the polarization matrix related
with the transitions between Mn($3d$) bands in order to get rid of the unphysical
metallic screening,
which is present in RPA if one starts from the LDA band structure \cite{Ferdi04}.

  Then, for each Mn-site, we obtain the
$5$$\times$$5$$\times$$5$$\times$$5$ matrix $\hat{U}^{\bf R}$,
which can generally depend on ${\bf R}$ and incorporate
some effects of the local environment in solids. The symmetry of these matrices
is also different from the spherical one.
Nevertheless, just for the explanatory purposes, we fit each matrix
in terms of three well known parameters, which would fully
specify all intraatomic interactions between $3d$ electrons
in the spherical environment: the Coulomb repulsion $U = F^0$,
the intraatomic exchange coupling $J = (F^2 + F^4)/14$, and the
``nonsphericity'' $B = (9F^2 - 5F^4)/441$,
where
$F^0$, $F^2$, and $F^4$ are radial Slater's integrals.
These parameters have the following meaning:
$U$ is responsible for the stability of certain atomic configuration
with the given number of electrons,
while $J$ and $B$ are responsible for the first and second Hund rule,
respectively. The results of such a fitting are shown in table \ref{tab:UJB}.
\Table{\label{tab:UJB}
Results of
parametrization of screened Coulomb interactions in terms of
the on-site Coulomb repulsion $U$, intraatomic exchange coupling $J$, and
``nonsphericity'' $B$ for the low-temperature ($T$$=$ 4 K) and high-temperature ($T$$=$ 550 K)
monoclinic phases of BiMnO$_3$. All parameters are measured in eV.
Positions of the Mn sites
are explained in figure \ref{fig.structure}.}
\br
& \centre{2}{$T$$=$ 4 K} & \centre{2}{$T$$=$ 550 K} \\
\ns
& \crule{2} & \crule{2} \\
interaction parameters & site 1 & site 3 & site 1 & site 3 \\
\mr
 $U$  & 2.27 & 2.27 & 2.31 & 2.22 \\
 $J$  & 0.89 & 0.88 & 0.89 & 0.88 \\
 $B$  & 0.09 & 0.09 & 0.09 & 0.09 \\
\br
\end{tabular}
\end{indented}
\end{table}
One can clear see that the Coulomb repulsion $U$ is greatly reduced due to the
self-screening effects, which are related with the admixture of the Mn($3d$)
states into the O($2p$) band.

\section{\label{sec:ModelAnalysis} Analysis of the Model Hamiltonian}

\subsection{\label{sec:HFApproximation}Hartree-Fock Approximation}

  In order to solve the model Hamiltonian (\ref{eqn:Hmanybody})
we employ the simplest mean-field Hartree-Fock approximation, where
the trial
many-electron
wavefunction is searched in the form of a single Slater determinant
$|S\{ \varphi_k^s \} \rangle$, constructed from the one-electron orbitals
$\{ \varphi_k^s \}$.
In this notation, $k$ is a collective
index combining the momentum
${\bf k}$ in the first Brillouin zone and the band number, and $s$ is
the spin of the particle.
The one-electron orbitals are requested to minimize
the total energy
\begin{equation}
E_{\rm HF}= \min_{\{ \varphi_k^s \}}
\langle S\{ \varphi_k^s \}|\hat{\cal H}| S\{ \varphi_k^s \} \rangle
\label{eqn:EHF}
\end{equation}
for a given number of particles $\cal{N}$.
The minimization is equivalent to the
solution of Hartree-Fock equations for $\{ \varphi_k^s \}$:
\begin{equation}
\left( \hat{t}_{\bf k} + \hat{\cal V}^s \right) | \varphi_{k}^s \rangle =
\varepsilon_k^s | \varphi_k^s \rangle,
\label{eqn:HFeq}
\end{equation}
where
$\hat{t}_{\bf k}$$\equiv$$\| t_{\bf k}^{m_1 m_2} \|$
is the one-electron part of the model Hamiltonian (\ref{eqn:Hmanybody}) in the reciprocal space,
$t_{\bf k}^{m_1 m_2}$$=
$$\sum_{{\bf R}'} t^{m_1 m_2}_{{\bf R}{\bf R}'} e^{-i {\bf k}  \cdot ({\bf R}-{\bf R}')}$,
and $\hat{\cal V}^s$$\equiv$$\| {\cal V}_{{\bf R} m_1 m_2}^s \|$ is the Hartree-Fock potential,
\begin{equation}
\fl
{\cal V}^\uparrow_{{\bf R} m_1 m_2} = \sum_{m_3 m_4}
\left\{ U^{\bf R}_{m_1 m_2 m_3 m_4} \left( n^\uparrow_{{\bf R}m_3 m_4} + n^\downarrow_{{\bf R}m_3 m_4} \right)
- U^{\bf R}_{m_1 m_4 m_3 m_2}n^\uparrow_{{\bf R}m_3 m_4} \right\}
\label{eqn:HFpot}
\end{equation}
(similar equation for ${\cal V}^\downarrow_{{\bf R} m_1 m_2}$ is obtained by interchanging
$\uparrow$ and $\downarrow$).
Equations (\ref{eqn:HFeq}) are solved self-consistently together with the equation
$$
\hat{n}^s = \sum_k^{occ} | \varphi_k^s \rangle \langle \varphi_k^s |
$$
for the
density matrix $\hat{n}^s \equiv \|n^s_{{\bf R} m_1 m_2}\|$ in the basis
of Wannier functions.

  After self-consistency, the total energy (\ref{eqn:EHF})
can be computed as
$$
E_{\rm HF} = \sum_{k s}^{occ} \varepsilon_k^s -\frac{1}{2}
\sum_{{\bf R} s} \sum_{m_1 m_2} {\cal V}^s_{{\bf R} m_2 m_1} n^s_{{\bf R} m_1 m_2}.
$$

\subsection{\label{sec:MagneticInteractions}Magnetic Interactions}

  By knowing $\{ \varepsilon_k \}$ and $\{ \varphi_k  \}$,
one can construct the one-electron (retarded) Green function,
$$
\hat{\cal G}^s_{{\bf RR}'}(\omega) = \sum_k
\frac{ | \varphi_k^s \rangle \langle \varphi_k^s |}
{ \omega - \varepsilon_k^s + i\delta } e^{i {\bf k}  \cdot  ({\bf R}-{\bf R}')},
$$
which can be used in many applications. For example, the interatomic magnetic
interactions corresponding to infinitesimal rotations of
the
spin magnetic moments near the equilibrium can be easily computed as \cite{Liechtenstein,TRN}:
\begin{equation}
J_{{\bf RR}'} = \frac{1}{2 \pi} {\rm Im} \int_{-\infty}^{\varepsilon_{\rm F}}
d \omega {\rm Tr}_L \left\{ \hat{\cal G}_{{\bf RR}'}^\uparrow (\omega)
\Delta \hat{\cal V}_{{\bf R}'} \hat{\cal G}_{{\bf R}'{\bf R}}^\downarrow (\omega)
\Delta \hat{\cal V}_{\bf R} \right\},
\label{eqn:JHeisenberg}
\end{equation}
where
$\Delta \hat{\cal V}_{\bf R} = \hat{\cal V}^\uparrow_{\bf R} - \hat{\cal V}^\downarrow_{\bf R}$
is the magnetic (spin) part of the Hartree-Fock potential,
${\rm Tr}_L$ denotes the trace over the orbital indices,
and
$\varepsilon_{\rm F}$ is the Fermi energy.
According to  (\ref{eqn:JHeisenberg}),
$J_{{\bf RR}'}$$>$$0$ ($<$$0$)
means that for the
given magnetic
state, the spin alignment in the bond $\langle {\bf RR}' \rangle$
corresponds to the local minimum (maximum) of the total energy.
However, in the following we will use the universal notations,
according to which $J_{{\bf RR}'}$$>$$0$ and $<$$0$ will
stand
the ferromagnetic and antiferromagnetic
coupling, respectively.
These notations correspond to the mapping of the total energy change
of the Hartree-Fock method,
associated with the small rotations of the magnetic moments,
onto the Heisenberg model \cite{Liechtenstein}:
$$
E_{\rm Heis} = -\frac{1}{2} \sum_{{\bf RR}'} J_{{\bf RR}'} {\bf e}_{\bf R} \cdot {\bf e}_{{\bf R}'},
$$
where ${\bf e}_{\bf R}$ is the direction of the spin magnetic moment at the
site ${\bf R}$.

  Generally, the parameters $\{ J_{{\bf RR}'} \}$ are not universal and
depend on the magnetic state in which they are calculated
(for example, through the change of the orbital ordering \cite{PRB06b}
or the electronic structure \cite{Springer} in each magnetic state).

  If we are dealing with the collinear magnetic structure,
where all spins are parallel to the $z$-axis, i.e.
${\bf e}_{\bf R}$$=$ ($0$,$0$,$1$) or ($0$,$0$,$-$$1$),
one can consider a small rotation of the magnetic moment at one of the site,
${\bf e}_{\bf R} = (\cos \theta_{\bf R} \sin \phi_{\bf R}, \sin \theta_{\bf R} \sin \phi_{\bf R}, \cos \theta_{\bf R})$,
and calculate the second derivative of $E_{\rm Heis}$ with respect to
$\theta_{\bf R}$:
\begin{equation}
J^0_{\bf R} = \sum_{{\bf R}'} s_{{\bf RR}'} J_{{\bf RR}'}.
\label{eqn:J0}
\end{equation}
In this expression,
$s_{{\bf RR}'}$$=$ $1$ and $-$$1$
stands correspondingly
for the FM and AFM alignment in the
bond $\langle {\bf RR}' \rangle$.
$J^0_{\bf R}$ characterizes the stability of the magnetic system
with respect to the rotation of the single spin. It can be also related with the
spin stiffness and the magnetic transition temperature in the mean-field
approximation \cite{Liechtenstein}.
If $J^0_{\bf R} > 0$, the spin system is stable while if
$J^0_{\bf R} < 0$ it is unstable.

\subsection{\label{sec:DESE}Decomposition into ``Double Exchange'' and ``Superexchange''}

  Many properties of perovskite manganese oxides are related with the
simple fact that the exchange splitting $\Delta \hat{\cal V}_{\bf R}$
is large, and for many applications can be treated as the largest
parameter in the problem \cite{Springer,PRL99}.
This is because Mn$^{3+}$
ions have four unpaired $3d$ electrons, which interact through the
Hund's rule coupling $J$. Loosely speaking, the exchange splitting between
the majority and minority spin states is controlled by the parameter $U$$+$$3J$,
which is about 4.9 eV (table \ref{tab:UJB}),
whereas the orbital polarization (or the splitting of occupied states
with one particular projection of spin) is controlled by $U$$-$$J$,
being ``only'' about 1.4 eV. Therefore, as the first approximation, one can
neglect the orbital dependence of $\Delta \hat{\cal V}_{\bf R}$
end replace it by some constant exchange splitting $\Delta_{\rm ex}$:
i.e.,
\begin{equation}
\Delta {\cal V}_{{\bf R} mm'} \rightarrow \Delta_{\rm ex} \delta_{mm'}.
\label{eqn:DEapproximation}
\end{equation}
A typical example of the exchange splitting in the low-temperature
monoclinic phase is shown in figure \ref{fig.ExchangeSplitting}:
\begin{figure}[h!]
\centering \noindent
\resizebox{8cm}{!}{\includegraphics{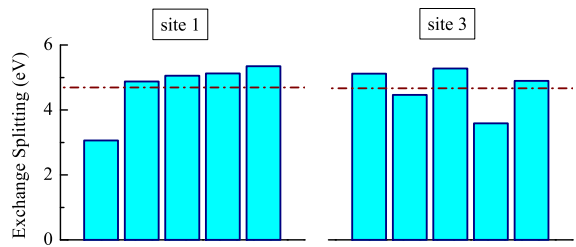}}
\caption{\label{fig.ExchangeSplitting}
Diagonal matrix elements of the exchange splitting for five $d$ orbitals
in the low-temperature monoclinic phase.
The off-diagonal matrix elements are considerably smaller.
The dash-dotted line shows the averaged value of the exchange splitting
(approximately 4.7 eV).
Positions of the Mn sites
are explained in figure \ref{fig.structure}.}
\end{figure}
the averaged exchange splitting $\Delta_{\rm ex}$ is about 4.7 eV,
whereas the deviations from $\Delta_{\rm ex}$ for the particular orbitals
do not exceed 1.5 eV. Of course, (\ref{eqn:DEapproximation}) is a crude
approximation. Nevertheless, as will see below, it appears to be very
useful for the analysis of interatomic magnetic interactions. It also reproduces
the main trends
of the behavior of these interactions
at least on the semi-quantitative level.

\begin{figure}[h!]
\centering \noindent
\resizebox{15cm}{!}{\includegraphics{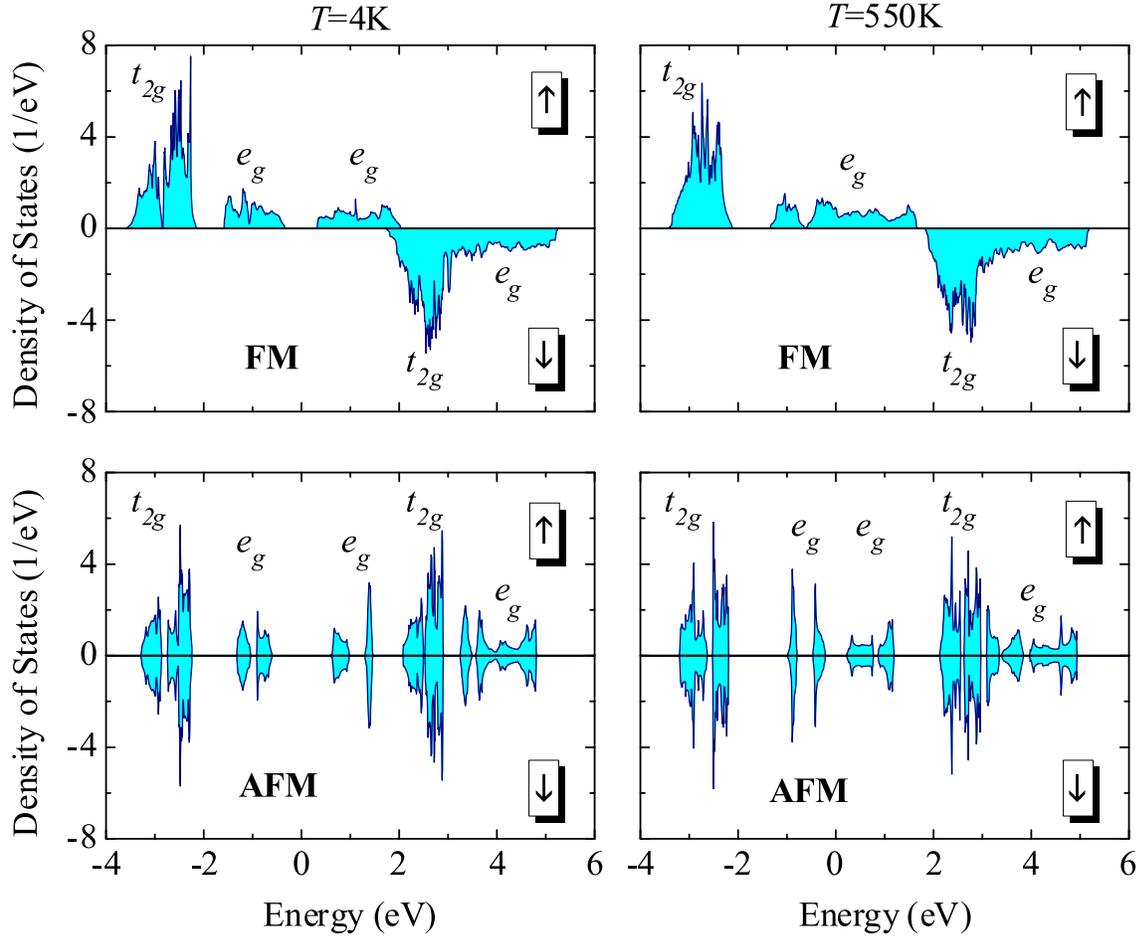}}
\caption{\label{fig.HFDOS}
Density of states in the
ferromagnetic (top) and antiferromagnetic
$\uparrow \downarrow \downarrow \uparrow$ (bottom)
phases of BiMnO$_3$, as obtained in the
Hartree-Fock calculations using low-temperature (left)
and high-temperature (right) monoclinic structures.
The Fermi level is at zero energy
(shown by dash-dotted line).
Symbols show positions of the main bands.}
\end{figure}

  Since $\Delta_{\rm ex}$ is large,
all minority-spin states are empty (figure \ref{fig.HFDOS}).
Therefore,
all poles of
$\hat{\cal G}_{{\bf R}'{\bf R}}^\downarrow$ are located in the
unoccupied part of the spectrum and below $\varepsilon_{\rm F}$
one can use the $1/\Delta_{\rm ex}$
expansion for $\hat{\cal G}_{{\bf R}'{\bf R}}^\downarrow$ \cite{Springer,PRL99}.
Then, the first two terms in the expansion of $J_{{\bf RR}'}$ will have the
following from:
\begin{equation}
J^D_{{\bf RR}'}=-\frac{1}{2\pi} {\rm Im} \int_{-\infty}^{\varepsilon_F}
d \omega {\rm Tr}_L \left\{
\hat{\cal G}^\uparrow_{{\bf RR}'}(\omega)
\hat{h}_{{\bf R}'{\bf R}}
 \right\},
\label{eqn:JDE}
\end{equation}
and
\begin{equation}
J^S_{{\bf RR}'}=-\frac{1}{2\pi \Delta_{\rm ex}} {\rm Im} \int_{-\infty}^{\varepsilon_F}
d \omega {\rm Tr}_L \left\{
\hat{\cal G}^\uparrow_{{\bf RR}'}(\omega)
\left(2\omega \hat{h}_{{\bf R}'{\bf R}} + (\hat{h})^2_{{\bf R}'{\bf R}}\right)\right\},
\label{eqn:JSE}
\end{equation}
where we have used the notations
$\hat{h}_{{\bf R}'{\bf R}} = \hat{t}_{{\bf R}'{\bf R}} + \hat{\cal V}^\uparrow_{\bf R}\delta_{{\bf R}'{\bf R}}$
and $(\hat{h})^2_{{\bf R}'{\bf R}} =
\sum_{{\bf R}''} \hat{h}_{{\bf R}'{\bf R}''}  \hat{h}_{{\bf R}''{\bf R}}$.

  $J^D_{{\bf RR}'}$ is proportional to
$\{ \hat{t}_{{\bf R}'{\bf R}} \}$ and does not depend on $\Delta_{\rm ex}$.
In an analogy with \cite{Springer,PRL99}, we will called it ``the double exchange interaction'',
although, strictly speaking, it is not a regular double exchange since
$\hat{\cal G}^\uparrow_{{\bf RR}'}$ also
includes $\hat{\cal V}^\uparrow_{\bf R}$, which
takes into accounts the effects of the orbital
polarization of the electronic origin. $J^S_{{\bf RR}'}$ incorporates the effects of the
second order with respect to $\{ \hat{t}_{{\bf R}'{\bf R}} \}$ and is inversely
proportional to $\Delta_{\rm ex}$. Therefore, in the following it will be called
``the superexchange interaction''. We will also consider two approximation for
$J^D_{{\bf RR}'}$ and $J^S_{{\bf RR}'}$. In the first one, $\hat{h}_{{\bf R}'{\bf R}}$
will be regular Hartree-Fock Hamiltonian for the majority-spin states and
$\hat{\cal G}^\uparrow_{{\bf RR}'}$ is the Green function corresponding to this
Hamiltonian:
\begin{equation}
\hat{\cal G}^\uparrow_{{\bf RR}'} = \left[\omega - \hat{h} + i\delta \right]^{-1}_{{\bf RR}'}.
\label{eqn:GF}
\end{equation}
In the second one,
in order to be consistent with the approximate expression (\ref{eqn:DEapproximation})
for the exchange splitting,
we will neglect all effects of the orbital polarization
of the electronic origin also in the definition of $\hat{h}_{{\bf R}'{\bf R}}$
and $\hat{\cal G}^\uparrow_{{\bf RR}'}$. Therefore, apart from the constant shift,
$\hat{h}_{{\bf R}'{\bf R}}$ is replaced by $\hat{t}_{{\bf R}'{\bf R}}$. Then,
$\hat{\cal G}^\uparrow_{{\bf RR}'}$ is the regular LDA Green function, which is
obtained from (\ref{eqn:GF}) after replacing $\hat{h}$ by $\hat{t}$.
In this approximation, it becomes more
clear why we continue to use the term ``double exchange'', even though our system
can be insulating, like the low-temperature phase of BiMnO$_3$, where
already in LDA
there is a
gap between Mn($e_g$) bands (figure \ref{fig.ek}).
It is true that the existence of this gap, $\Delta$, is related with some kind
of the orbital polarization. In this sense
it is still reasonable to consider the superexchange
processes by treating all transfer integrals as a perturbation.
This
would correspond to the
superexchange interactions of
the form $t_{\rm eff}^2/\Delta$, where $t^2_{\rm eff}$ is
the square of an effective
transfer integral between Mn sites, which is related with $\{ \hat{t}_{{\bf R}'{\bf R}} \}$.
However,
this orbital polarization comes from the large crystal-field splitting,
which is just another effect of the covalent mixing, and, therefore
has the same origin as $t_{\rm eff}$. Therefore, $\Delta$ should be proportional to
$t_{\rm eff}$, and the ``superexchange'' $t_{\rm eff}^2/\Delta$ becomes also proportional
to $t_{\rm eff}$. From this point of view, it is still reasonable to call this interaction
as the ``double exchange''.

\section{\label{sec:Results} Results and Discussion}

\begin{figure}[h!]
\centering \noindent
\resizebox{5cm}{!}{\includegraphics{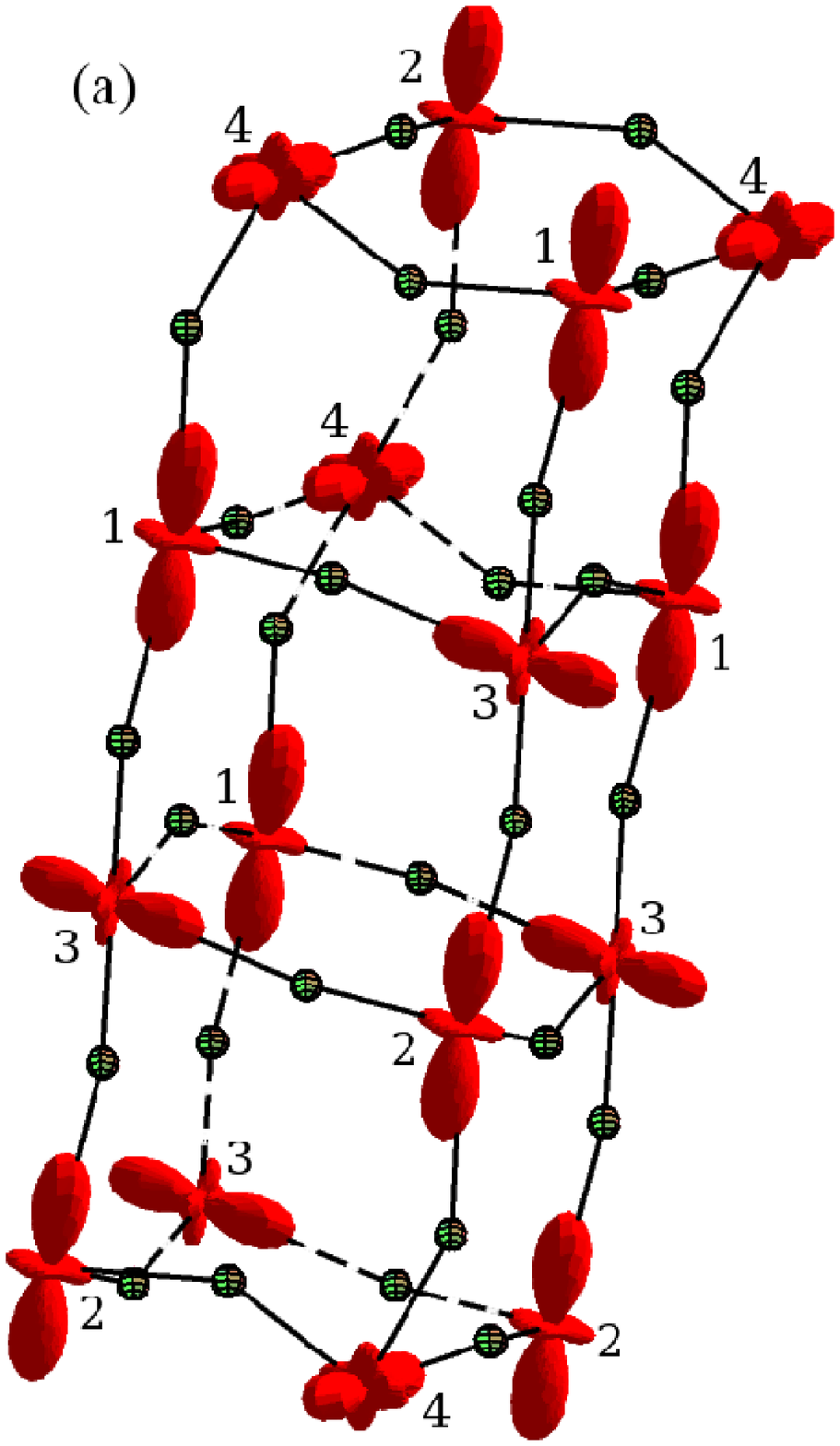}}
\resizebox{5cm}{!}{\includegraphics{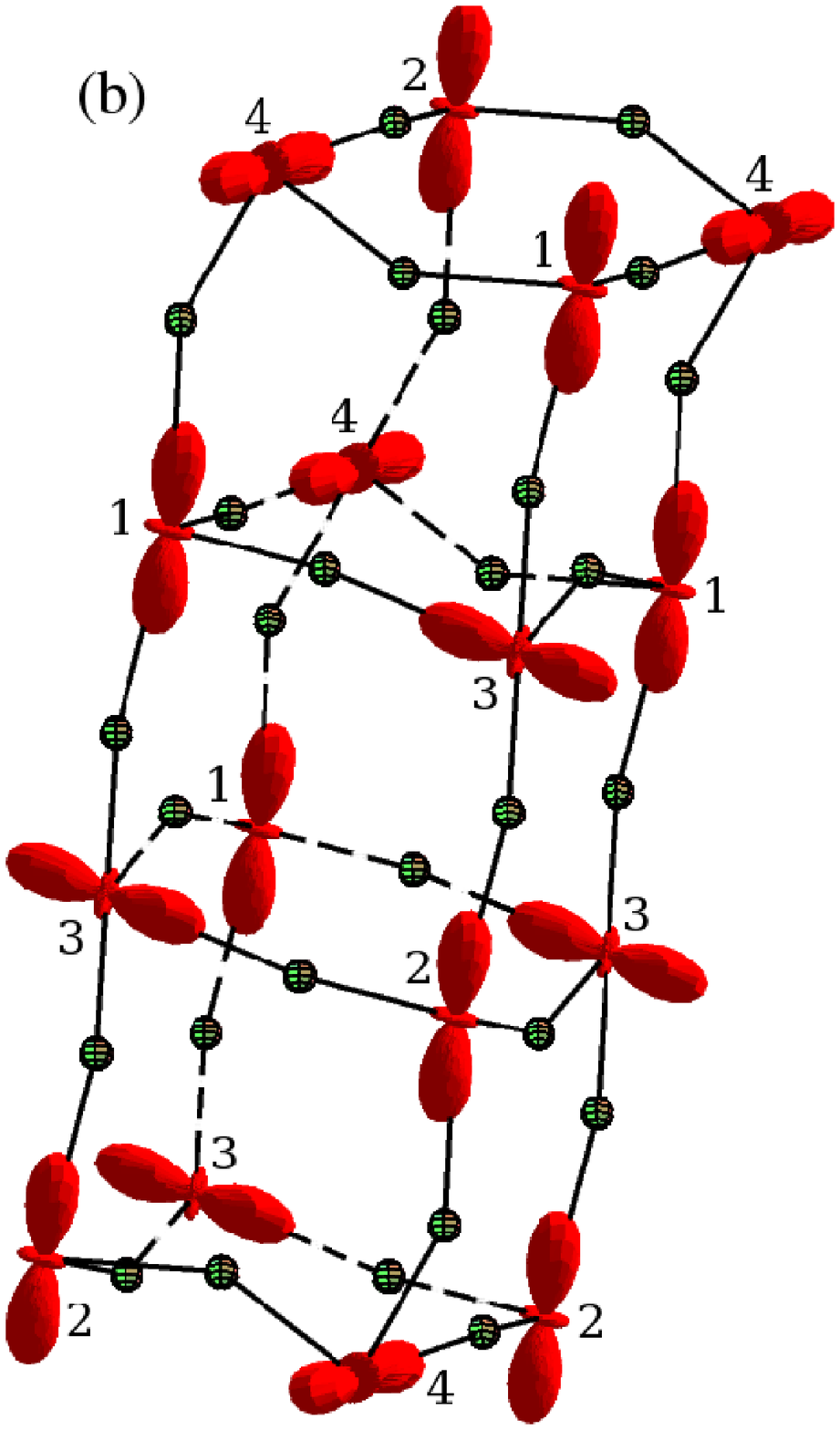}}
\resizebox{5cm}{!}{\includegraphics{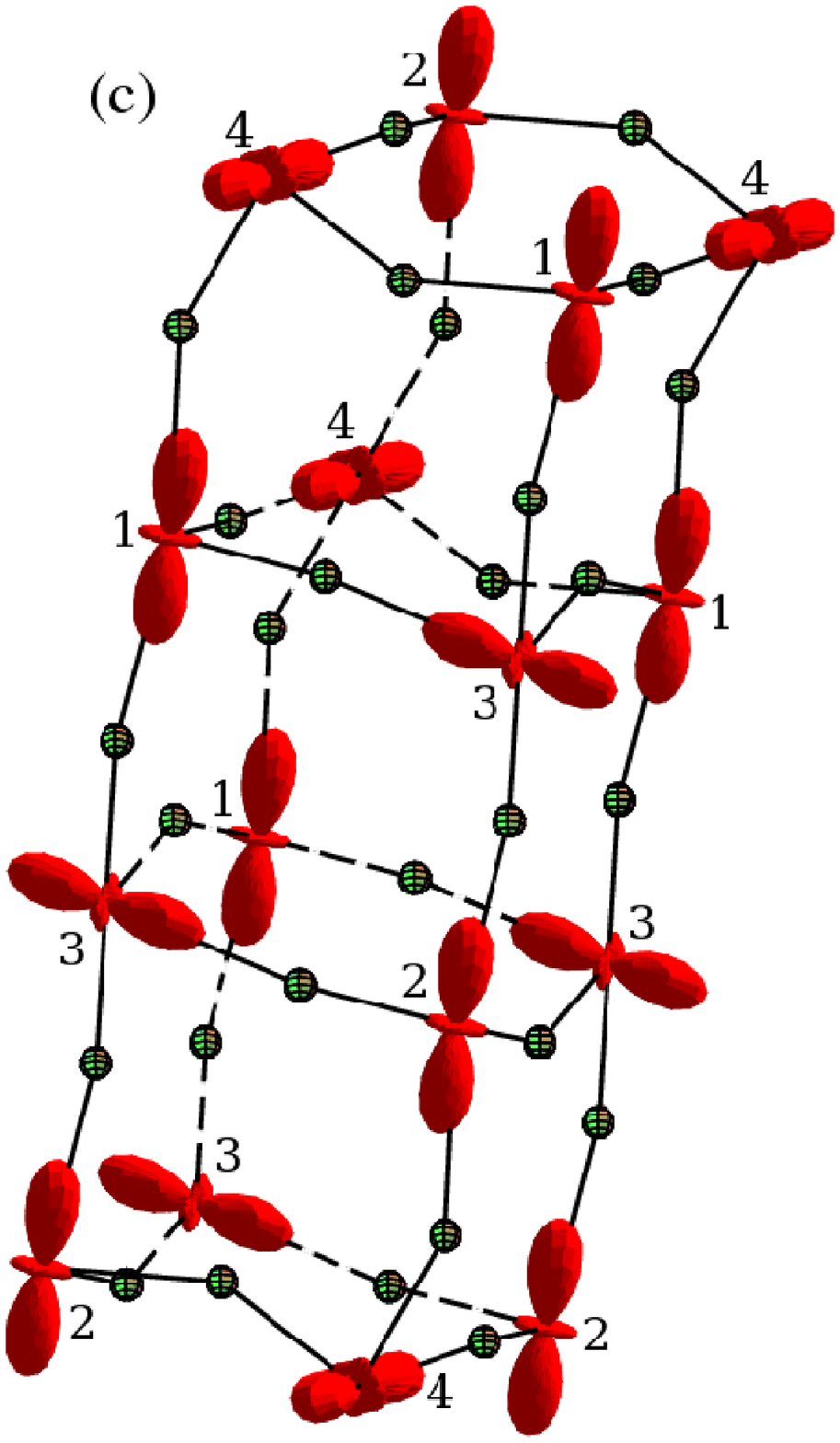}}
\caption{\label{fig.OrbitalOrdering4K}
Orbital ordering in the low-temperature monoclinic phase of BiMnO$_3$
obtained through the site-diagonal elements of the density matrix in the original LMTO basis (a),
through the crystal-field orbitals of downfolded Hamiltonian (b),
and through the density matrix derived in Hartree-Fock
calculations for the ferromagnetic state (c).
Four types of Mn atoms are indicated by the numbers.}
\end{figure}

   The orbital ordering in the low-temperature monoclinic phase of BiMnO$_3$ is shown
in figure \ref{fig.OrbitalOrdering4K}.
We have tried three different methods in order to derive the
distribution of the electron density around Mn-sites.
\begin{enumerate}
\item
In the first method, we simply calculate the site-diagonal elements
of the density matrix in the original LMTO basis
by integrating over four lowest $e_g$ bands (spreading around 1 eV in figure \ref{fig.ek}),
and plot the electron density corresponding to this density matrix.
\item
In the second method,
we plot the densities of the lowest $e_g$-orbitals obtained from the diagonalization
of the site-diagonal part of the one-electron Hamiltonian, which was derived from the
downfolding method. In the other words, these are just the crystal-field orbitals, corresponding
to the fourth atomic level in figure \ref{fig.CF}.
\item
In the third method,
we plot the electron density for the occupied $\uparrow$-spin $e_g$ band, obtained in the
Hartree-Fock calculations for the ferromagnetic state (figure \ref{fig.HFDOS}).
\end{enumerate}
All three methods provide a very consistent picture for the
general details of the orbital ordering in
the low-temperature phase of BiMnO$_3$,
which is also consistent with results of full-potential calculations by Shishidou \cite{Shishidou}.

\begin{figure}[h!]
\centering \noindent
\resizebox{8cm}{!}{\includegraphics{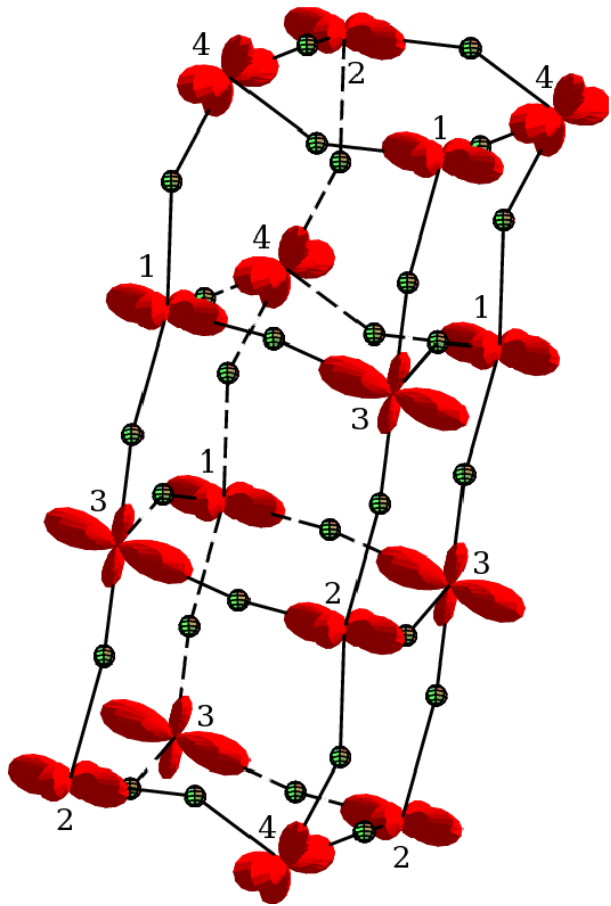}}
\caption{\label{fig.OrbitalOrdering550K}
Orbital ordering in the high-temperature monoclinic structure of BiMnO$_3$.
Four types of Mn atoms are indicated by the numbers.}
\end{figure}

  The orbital ordering in the high-temperature structure, derived from the crystal-field
orbitals, is shown in figure \ref{fig.OrbitalOrdering550K}.
This orbital ordering is entirely related with the crystal distortion, which splits the atomic
$e_g$ levels. Even for the sites `1' and `2' with the least distorted environment (table \ref{fig.CF}),
this splitting is of the order of 0.2 eV. This corresponds to the temperature
of about 2100 K, which largely exceed the
temperature of monoclinic-to-orthorhombic transition (about 770 K \cite{Kimura}).
Thus, it is reasonable to expect the orbital ordering to take place in \textit{both}
monoclinic phases, below and above 474 K. However, as it is clearly seen from the
comparison of figures \ref{fig.OrbitalOrdering4K} and \ref{fig.OrbitalOrdering550K},
the character of the orbital ordering will change at the point of phase transition.
This conclusion
is qualitatively consistent with results of resonant x-ray scattering
on BiMnO$_3$ \cite{Yang}.

  Results of Hartree-Fock calculations of the total energies for the
ferromagnetic and several antiferromagnetic configurations are shown
in table \ref{tab:TotalEnergies}.
\begin{table}[h!]
\caption{Total energies
for several antiferromagnetic configurations
as obtained in the Hartree-Fock calculations for the low-temperature
($T$$=$ 4 K) and high-temperature ($T$$=$ 550 K) monoclinic structures of BiMnO$_3$.
The energies measured in meV per one formula unit
relative to the ferromagnetic state.}
\label{tab:TotalEnergies}
\begin{indented}
\item[]\begin{tabular}{@{}ccc}
\br
 configuration & $T$$=$ 4 K & $T$$=$ 550 K \\
\mr
 $\uparrow \downarrow \downarrow \uparrow$       & -0.5   & \phantom{2}9.4   \\
 $\uparrow \uparrow   \downarrow \downarrow$     & 19.0   & 48.4   \\
 $\uparrow \downarrow \downarrow \downarrow$     & \phantom{1}3.9   & 28.4   \\
 $\downarrow \downarrow \uparrow \downarrow$     & \phantom{1}5.6   & 21.5   \\
\br
\end{tabular}
\end{indented}
\end{table}
Our main observation is that for the low-temperature
monoclinic structure the ferromagnetic phase appears to be nearly
degenerate with the antiferromagnetic $\uparrow \downarrow \downarrow \uparrow$
phase, which can be obtained by flipping
the directions of the magnetic moments at the Mn-sites `2' and `3'.
Perhaps, the tendencies towards the antiferromagnetism are somewhat overestimated
in our model, and there are several reasons for it:
\begin{enumerate}
\item
The calculations of the Coulomb interaction $U$ are
always conjugated with certain approximations \cite{PRB06a,review2008},
and some of these parameters may be underestimated. As we will see below,
larger values of the parameter $U$ would indeed help in
stabilizing the ferromagnetic phase.
\item
Our model (\ref{eqn:Hmanybody}) does not explicitly include the oxygen states.
This appears to be a good approximation for titanium and
vanadium perovskite oxides \cite{PRB06b}, where the transition-metal and oxygen bands
are well separated. However,
manganese compounds are much closer to the charge-transfer
regime because of the proximity of Mn($3d$) and O($2p$) bands, and
much stronger hybridization, which takes place between these
groups of states. Moreover,
as it was pointed out in section \ref{sec:OneElectron},
there is also
an overlap between Mn($3d$) and Bi($6p$) bands. Therefore, the
Hubbard
model (\ref{eqn:Hmanybody}),
where the form of the Coulomb interactions is borrowed
from the atomic limit for the Mn($3d$) states is an approximation, which may
ignore some contributions to the relative stability of different magnetic configurations.
For example, it is known that the magnetic polarization of the oxygen states
will additionally stabilize the FM phase \cite{Mazurenko}.
These effects are not included into the model (\ref{eqn:Hmanybody}).
\end{enumerate}
Nevertheless, as we will see below,
the competition between ferromagnetic
and antiferromagnetic $\uparrow \downarrow \downarrow \uparrow$ phases
itself is a genuine effect, which is directly related with the form of
the orbital ordering in the low-temperature monoclinic structure.

\begin{figure}[h!]
\centering \noindent
\resizebox{8cm}{!}{\includegraphics{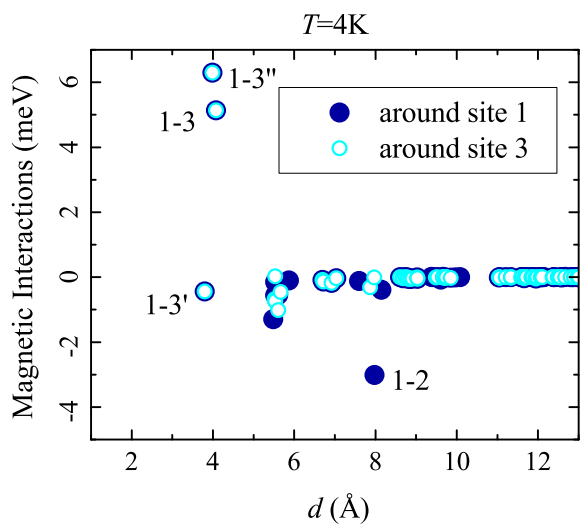}}
\caption{\label{fig.exchange}
Distance-dependence of interatomic magnetic interactions
in the low-temperature monoclinic structure of BiMnO$_3$.
The values around two inequivalent
Mn sites are shown by closed and open symbols.
Other notations indicate the bonds
for the most relevant magnetic interactions.
Positions of the Mn sites
are explained in figures \protect\ref{fig.structure} and \protect\ref{fig.OrbitalOrdering}.}
\end{figure}

\begin{figure}[h!]
\centering \noindent
\resizebox{8cm}{!}{\includegraphics{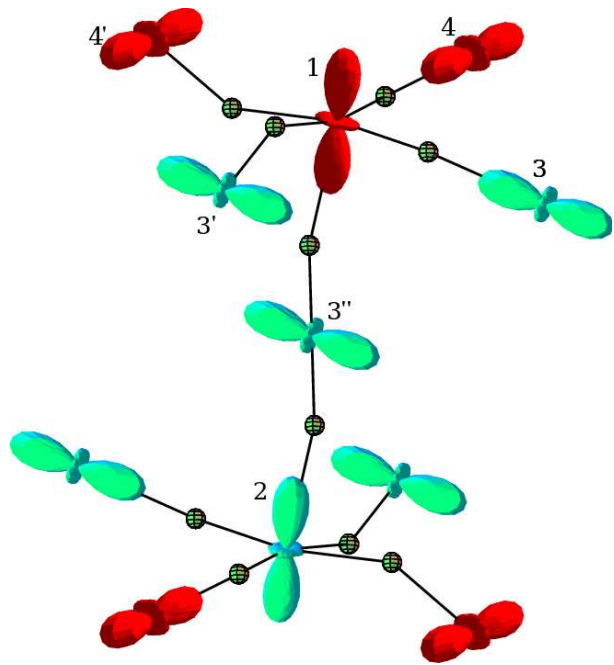}}
\caption{\label{fig.OrbitalOrdering}
Fragment of the orbital ordering pattern in the low-temperature
monoclinic phase of
BiMnO$_3$. Different magnetic sublattices, which are formed in the
antiferromagnetic
$\uparrow \downarrow \downarrow \uparrow$ structure
are shown by different colors.}
\end{figure}

  The distance-dependence of interatomic magnetic interactions calculated
in the low-temperature monoclinic
structure is shown in figure \ref{fig.exchange}.
These calculations have been performed using
the formula (\ref{eqn:JHeisenberg}) for infinitesimal rotations
of magnetic moments near the ferromagnetic state.
We note the following. There are two types of relatively strong ferromagnetic
interactions between nearest neighbors, which operate in the bonds
$1$-$3$ and $1$-$3''$ (see figure \ref{fig.OrbitalOrdering}
for notations).\footnote{
Since in the ferromagnetic structure, the atoms $3$ and $4$ can be
transformed to each other by the symmetry operations, similar interactions hold
between atoms $1$ and $4$. However, for the sake of clarity, we do not show all
these bonds in figure \ref{fig.OrbitalOrdering}.}
The character of these interactions is directly related with the
``antiferromagnetic'' orbital ordering in the bonds $1$-$3$ and $1$-$3''$, and can
be anticipated already from
distribution of the Mn-O bondlengths in the low-temperature monoclinic
phase \cite{belik_07}.\footnote{
In the present context, the ``antiferromagnetic orbital ordering'' means
nearly orthogonal orientation of occupied $e_g$ orbitals, which maximizes
the electron hoppings into the unoccupied subspace and, therefore,
favors the ferromagnetic interactions
between these sites \cite{KugelKhomskii}.}
The interaction in the third bond $1$-$3'$, formed by the nearest neighbors,
is relatively week. This is again consistent with the geometry of the
orbital ordering, corresponding to the minimal overlap between
occupied and unoccupied
$e_g$ orbitals. Similar situation occurs in the bond
$1$-$4'$.
The most striking result of the present calculations is the
existence of relatively strong
long-range
antiferromagnetic interaction in the bond $1$-$2$ (figure \ref{fig.OrbitalOrdering}).
Nevertheless, this result is also anticipated from the geometry of the orbital ordering.
Note that the occupied $e_g$ orbitals at the sites $1$ and $2$ are directed
towards each other. Although the direct transfer integrals
between these two sites are relatively small
(see figure \ref{fig.transfer}),
it is still reasonable to expect the existence of AFM interactions,
which are mediated by unoccupied $e_g$ states of the site $3''$.
Such a situation is somewhat similar to superexchange interactions,
which take place via oxygen states in the
charge-transfer insulators \cite{Oguchi,ZaanenSawatzky}, and the mechanism itself
is sometimes called as the ``super-superexchange''.

  Thus, in the low-temperature monoclinic phase of
BiMnO$_3$ we are always dealing with a competition of
nearest-neighbor FM and longer-range AFM interactions.
In fact,
there are several factors, which can make these interactions comparable with each other.
It is true that, generally, the nearest-neighbor interactions are expected to be much stronger,
because all transfer integrals are basically restricted by the nearest neighbors
(figure \ref{fig.transfer}). However, for the nearest-neighbor interactions we are
also dealing with the strong cancelation of FM ``double exchange''
and AFM superexchange contributions (table \ref{tab:MInteractions}).
\Table{\label{tab:MInteractions}
Magnetic interactions in ferromagnetic (FM) and antiferromagnetic (AFM)
$\uparrow \downarrow \downarrow \uparrow$
phases as obtained for the low-temperature ($T$$=$ 4 K) and high-temperature ($T$$=$ 550 K)
monoclinic structures.
The columns `DE' and `SE' show results of (approximate) decomposition into double exchange and
superexchange contribution using electronic structure obtained in the Hartree-Fock
calculations for the ferromagnetic state and the one in LDA (in the parenthesis).
All values are in meV. Positions of Mn sites are explained in figure \ref{fig.OrbitalOrdering}.}
\br
& \centre{4}{$T$$=$ 4 K} & \centre{3}{$T$$=$ 550 K} \\
\ns
& \crule{4} & \crule{3} \\
\br
 bond       & \centre{1}{FM}     & \centre{1}{AFM}   & \centre{1}{DE}                   & \centre{1}{SE} &
 \centre{1}{FM}     & \centre{1}{DE}        & \centre{1}{SE}      \\
\mr
 $1$-$3$    &  \phantom{-}5.1   &  \phantom{-}1.7   & \phantom{-}32.9 (\phantom{-}42.0) &           -30.1 (-29.5) &
 15.9           &           43.2  &  -30.3  \\
 $1$-$4$    &  \phantom{-}5.1   &  \phantom{-}7.8   & \phantom{-}32.9 (\phantom{-}42.0) &           -30.1 (-29.5) &
 15.9           &           43.2  &  -30.0  \\
 $1$-$3'$   & -0.4              & -1.0              & \phantom{-}16.5 (\phantom{-}22.4) &           -20.4 (-21.7) &
 19.2           &           40.6  &  -24.0  \\
 $1$-$4'$   & -0.4              &  \phantom{-}0.7   & \phantom{-}16.5 (\phantom{-}22.4) &           -20.4 (-21.7) &
 19.2           &           40.6  &  -24.0  \\
 $1$-$3''$  &  \phantom{-}6.3   &  \phantom{-}5.1   & \phantom{-}29.8 (\phantom{-}36.8) &           -26.3 (-26.0) &
 18.5           &           41.7  &  -26.2  \\
 $2$-$3''$  &  \phantom{-}6.3   &  \phantom{-}5.6   & \phantom{-}29.8 (\phantom{-}36.8) &           -26.3 (-26.0) &
 18.5           &           41.7  &  -26.2  \\
 $1$-$2$    & -3.0              & -3.0              & \phantom{2}-0.5 (\phantom{2}-0.7) & \phantom{2}-1.4 (\phantom{2}-1.7) &
 \phantom{5}1.0 & \phantom{5}1.4  & \phantom{-2}0.2  \\
\br
\end{tabular}
\end{indented}
\end{table}
For example, this cancellation is nearly perfect for the ``weak bonds''
$1$-$3'$ and $1$-$4'$. This is a general rule for perovskite manganese
oxides, which explains a strong reduction of nearest-neighbor magnetic interactions,
so that
they can easily become comparable with some
longer-range interactions \cite{TRN,Springer,PRL99}.
On the other hands, for the longer-range AFM interaction
in the bond $1$-$2$, there is no such cancellation.
The long-range interactions are expected to vanish for the undoped (parent)
manganites,
provided that they would have a undistorted cubic structure.
This effect is entirely related with the symmetric filling (or half-filling)
of the majority-spin $e_g$ band \cite{PRL99}. Nevertheless, many
parent manganites (like BiMnO$_3$) have a strongly distorted crystal structure.
This distortion gives rise to the
orbital ordering, which leads to certain asymmetry of filling
of the majority-spin $e_g$ band, and this asymmetry is finally manifested
in the appearance of longer-range interactions.

  Since nearest-neighbor interactions favor the ferromagnetism,
while the longer-range interactions favor the formation of the
antiferromagnetic $\uparrow \downarrow \downarrow \uparrow$ structure,
one can generally expect the competition between these two phases, as
it is clearly seen from
results of total energy calculations shown in table \ref{tab:TotalEnergies}.
Nevertheless, there is another factor, which will additionally stabilizes
the AFM $\uparrow \downarrow \downarrow \uparrow$ phase
and explains why it is so close in energy to the
FM phase.

\begin{figure}[h!]
\centering \noindent
\resizebox{6cm}{!}{\includegraphics{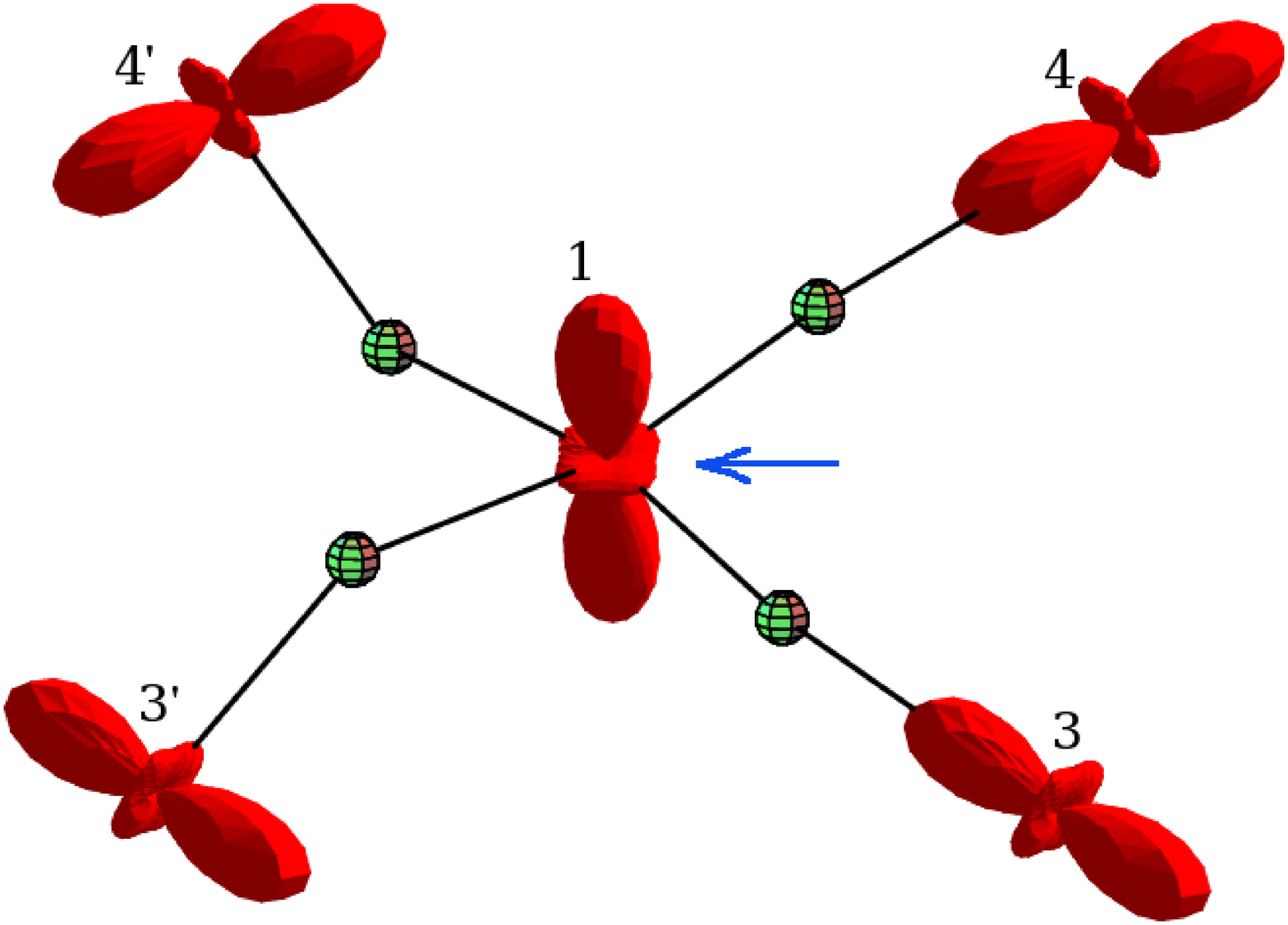}}
\resizebox{6cm}{!}{\includegraphics{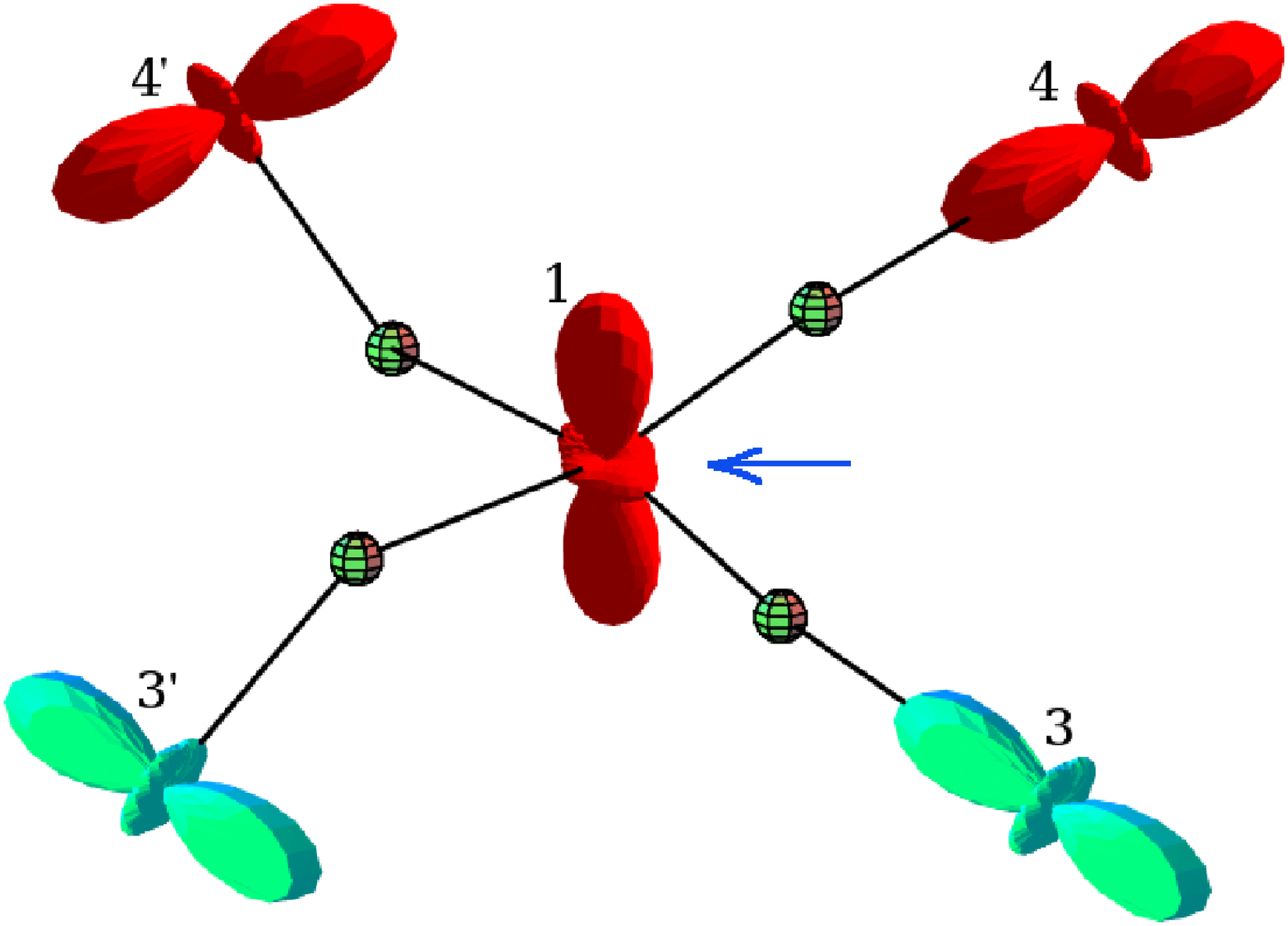}}
\caption{\label{fig.OrbitalOrdering4KDetails}
Details of orbital ordering in the ferromagnetic (left)
and antiferromagnetic (right) phases
realized in the low-temperature monoclinic structure of
BiMnO$_3$. Different magnetic sublattices are shown by different colors.
The arrow shows the region where the change of the orbital cloud results in the
drastic
change of interatomic magnetic interactions.}
\end{figure}

  Note that the orbital degrees of freedom can additionally
change their form in order to modify the interatomic magnetic interactions
in the direction which will further
stabilize the given magnetic structure \cite{KugelKhomskii}.
Since the form of the orbital ordering is efficiently
constrained by the large crystal-field splitting, it does
not strongly depend on the magnetic state, and visually one can
observe only tiny changes
in the distribution of occupied $e_g$ electron density
(figure \ref{fig.OrbitalOrdering4KDetails}).
Nevertheless,
it is interesting to see that these tiny changes may have a profound effect
on the behavior of interatomic magnetic interactions.
Indeed, in the AFM
$\uparrow \downarrow \downarrow \uparrow$ structure,
the chain $1$-$3''$-$2$ contains one AFM bond ($1$-$3''$) and
one FM bond ($2$-$3''$, see figure \ref{fig.OrbitalOrdering}).
Although in the AFM structure, both interactions remain ferromagnetic,
there is a clear \textit{polarity} of interactions and the magnetic
coupling in the AFM bond $1$-$3''$ is considerably weaker than
the one
in the ferromagnetic bond $2$-$3''$
(table \ref{tab:MInteractions}). Even more dramatic change
occurs in the plane of the distorted perovskite structure.
In the AFM structure,
even visually one can see some anisotropy in
distribution of the occupied $e_g$
electron density at the site $1$ (figure \ref{fig.OrbitalOrdering4KDetails}), which
appears to be more contracted in the direction of the
FM bond $1$-$4$.
On the other hand, this distribution is nearly isotropic
in the FM phase.
This means that
in the direction $1$-$4$ of the AFM phase,
the weight of the $e_g$ orbitals is additionally
moved into the unoccupied part of the spectrum.
This opens some additional pathes for
the virtual hoppings into the unoccupied part of the spectrum, which will
additionally
stabilize the
FM coupling.
Indeed, the exchange coupling in the FM bond $1$-$4$ is 7.8 meV,
while the one in the AFM bond $1$-$3$ is strongly reduced till
1.7 meV.
Moreover, the magnetic interactions in the weak bonds
$1$-$3'$ and $1$-$4'$ are also adjusted by the orbital-ordering effects.
For example, the AFM coupling in the bond $1$-$3'$ is enhanced, while
the coupling in the FM bond $1$-$4'$ becomes ferromagnetic.

  Because of these orbital ordering effects
both magnetic configurations appear to be locally stable.
Indeed, the parameters $J^0_{\bf R}$, calculated in the
FM and AFM $\uparrow \downarrow \downarrow \uparrow$ phases are
($7.4$, $12.1$) and ($17.6$, $8.9$) meV, respectively, where the
first number in parenthesis corresponds to the Mn-site 1, while
the second number corresponds to the Mn-site 3.
All parameters are positive, meaning that both configurations are stable,
at least with respect to independent rotations of magnetic
moments at the sites $1$ and $3$.

  In addition to electronic degrees of freedom,
the AFM $\uparrow \downarrow \downarrow \uparrow$
phase can be additionally stabilized by structural effects
associated with polar atomic displacements in the direction
which further minimizes the total energy of the system
via magneto-elastic interactions. For example, magnetic
couplings in the bonds $1$-$3''$ and $2$-$3''$
can be further adjusted by displacement of the Mn-atom $3''$
(figure \ref{fig.OrbitalOrdering}). Although we do not
consider such a mechanism here (that would require detailed
structural optimization using modern full-potential electronic
structure methods), it would be certainly an interesting
step to do in the future.
This mechanism was considered for example in \cite{Wang,Picozzi} for other
multiferroic compounds.

  In the high-temperature monoclinic structure, the FM phase
is clearly the most stable one (table \ref{tab:TotalEnergies}).
This is closely related with the fact that
the FM phase is metallic (figure \ref{fig.HFDOS})
and the double
exchange interactions clearly dominate (table \ref{tab:MInteractions}).
However, this is true only for
the low-temperature regime. At elevated temperatures, the structure of
interatomic
magnetic interactions will be largely modified by the
magnetic disorder, which may also destroy metallic
character of the electronic structure.\footnote{
In fact, the Hartree-Fock calculations for the high-temperature monoclinic
structure shown that only the FM phase is metallic, whereas
all considered AFM phases are insulating
(figure \ref{fig.HFDOS}).
}
Thus, it is rather meaningless to discuss the properties of the
high-temperature phase in the low-temperature limit. The magnetic disorder in the
high-temperature phase is certainly
one of the interesting problems. However,
it is beyond the scopes of the present work. The possible tools
to address this problem is the coherent potential approximation (CPA) \cite{deBrito,PRB03},
or, more generally, the dynamical mean-field theory (DMFT) \cite{DMFT}.
Nevertheless, we would like to emphasize that since the type of the
orbital ordering changes above 474 K (figure \ref{fig.OrbitalOrdering550K}),
the long-range interactions between atoms 1 and 2 are expected
to be weak. Therefore, the long-range AFM correlations,
which according to our point of view are indispensable for the
inversion symmetry breaking and
appearance of the ferroelectricity,
should not play any important role in the
high-temperature monoclinic phase.
Thus, according to our scenario, the temperature of the isostructural
phase transition (474 K) should be also regarded as a upper bound
for the possible onset of ferroelectricity.

\section{\label{sec:Summary} Implications to the Properties of BiMnO$_3$}

  Thus, we would like to propose that the multiferroic behavior of BiMnO$_3$
should be closely related with a competition between two magnetic phases.
One is the centrosymmetric (and antiferroelectric) FM phase.
Another one is the AFM
$\uparrow \downarrow \downarrow \uparrow$ phase, which breaks
the inversion symmetry and allows for the spontaneous electric polarization
in the direction perpendicular to the $y$-axis.
The existence of both phases is closely related with the peculiar
orbital ordering, which takes place below 474 K.

  This means that despite the fact that BiMnO$_3$ is crystallized in the
centrosymmetric $C2/c$ structure, there is still a room for the multiferroic behavior,
if we could engineer the samples where these two phases coexist in a narrow
energy range accessible for the physical changes of electric and magnetic
fields as well as the temperature $T$. Then, one could readily expect the
``switching phenomena''. For example, by applying the magnetic field $H$
one could stabilize the FM antiferroelectric phase and switch off the
net electric
polarization. Conversely, one could apply the electric field $E$ and
stabilize the AFM ferroelectric phase with zero net magnetization.

  It is true
the low-temperature ferromagnetism in the pure bulk samples of BiMnO$_3$
is well established
today \cite{belik_07}. Therefore, the symmetry is expected to be
$C2/c$ and these samples are not
extremely promising from the viewpoint of multiferroic applications.
However, it is also known that the BiMnO$_3$ is extremely difficult
to synthesize, especially in the single crystalline form.
Therefore,
it is reasonable to expect that the BiMnO$_3$ samples will always have some
defects, and we would like to speculate that these defects may play a
positive role in stabilizing some fractions of
antiferromagnetic and noncentrosymmetric
phase. This seems to be reasonable, because the
low-temperature magnetization observed in the BiMnO$_3$ samples,
which do exhibit the ferroelectric behavior, was only $2.6\mu_B$
and reached $3.1\mu_B$ per one formula unit in high magnetic fields \cite{SantosSSC}.
These values are considerably lower that $4\mu_B$ expected for
the single saturated FM phase, meaning that the samples were not sufficiently pure
and might contain a fraction of the AFM phase.
This strongly reminds the idea of clustering or
macroscopic phase separation, which
has been intensively discussed
in other perovskite manganese oxides,
in the context of
colossal magnetoresistance phenomena \cite{Nagaev,Dagotto}.

  In this respect it is important to note that
even in the high-quality samples, whose low-temperature
saturation magnetization was close to $4\mu_B$ \cite{belik_07},
the authors of \cite{Yokosawa} and \cite{Kodama},
by means of atomic
pair distribution function analysis on neutron powder
diffraction data and selected-area electron
diffraction technique, respectively,
have observed the existence of short-range
ordered structures (or domains)
with the broken inversion symmetry.
The symmetry of these structures was
either $P2$ or $P2_1$, which is consistent with the
crystal
symmetry of the AFM $\uparrow \downarrow \downarrow \uparrow$
phase ($P2$), that we propose.
Moreover, these experimental data strongly suggests that
the breaking of the inversion symmetry is mainly caused
by the Mn atoms, that is again consistent with the
idea of the magnetic origin of this effect.
Unfortunately, the authors of \cite{Yokosawa,Kodama}
focused only on the structural properties of BiMnO$_3$,
and did not provide any information on how
these structural properties can be related with
the magnetic behavior
of BiMnO$_3$.
From our point of view, such measurements would be
very useful. For example, if the origin of the
noncentrosymmetric domains was indeed magnetic, it is
reasonable to expect
the size and
relative wight
of these domains
to decrease
in the magnetic field.

  Another possibility to control the properties of BiMnO$_3$
is to use the thin films. In this case,
due to the lattice mismatch of the bulk BiMnO$_3$ and the
substrate, the latter causes an additional strain and may
strongly affect the magnetic behavior of BiMnO$_3$.
This effect is well known for other (colossal magnetoresistive)
manganese oxides, and can be used as an efficient tool for
controlling the electronic and magnetic properties of
these systems near the point of
phase transition between FM and AFM states \cite{Konishi,Fang}.
Moreover, the magnetic structure at the surface of BiMnO$_3$ may be
also different from
the one in the bulk.
Indeed, the saturation magnetization of the BiMnO$_3$ thin
films grown on the $(100)$ SrTiO$_3$ substrate
was only $2.8\mu_B$ \cite{Ohshima}, which is considerably smaller than the
bulk value.
The magnetic moment increases almost linearly
with the increases of the film thickness and
reaches the nearly saturated value at around 500 \AA. The use of the
$(110)$ SrTiO$_3$ substrate yields even smaller saturation moment
(about $1.8 \mu_B$ \cite{Ohshima}).
Thus, all these data suggest that the magnetic ground state realized in the
thin films of BiMnO$_3$ is not a pure FM one,
and may contain some elements of the AFM structure.
Of course, on the basis of this highly limited experimental information
about the saturation magnetization it is impossible to
make a ultimate conclusion whether
this AFM structure is indeed the
$\uparrow \downarrow \downarrow \uparrow$ one, which we propose.
Nevertheless, one can speculate that the ferroelectric behavior,
which is apparently seen in the BiMnO$_3$ thin films, can be again
related with the deviation of the magnetic structure from the
pure
FM one \cite{SantosSSC,Sharan}.

  Finally, although the ground state of BiMnO$_3$ is ferromagnetic,
some fraction of
the AFM $\uparrow \downarrow \downarrow \uparrow$ phase
can emerge at elevated temperatures.
If at zero temperature both FM and $\uparrow \downarrow \downarrow \uparrow$ AFM phases
correspond to the local minima of the total energy and are
connected to each other
by the first-order transition, the rising of the temperature can
naturally lead to a
coexistence of these two states.
By neglecting the interaction between two phases, this effects is simply related
with the configuration mixing entropy. Moreover, the appearance of the
two-phase state is considerably facilitated in the presence of defects \cite{Nagaev,Dagotto},
as it is well known for other (colossal magnetoresistive) manganites \cite{Uehara}.
This behavior implies the existence of the temperature hysteresis
loop
in the magnetization curve near $T_C$.
A small temperature hysteresis has been indeed reported in \cite{belik_07},
which may be related with small anomalies of dielectric constant
observed in \cite{Kimura}.
Since the existence of the long-range AFM interaction is closely related with the
peculiar orbital ordering
persisting up to 474 K, no ferroelectricity
can be generally expected above this temperature.

  In summary, we believe that further exploration of multiferroic behavior
of BiMnO$_3$ should be focused on the revealing of the AFM $\uparrow \downarrow \downarrow \uparrow$
phase.
Possible multiferroic applications of BiMnO$_3$ will strongly depend on whether one can find the
conditions of coexistence of the FM and AFM $\uparrow \downarrow \downarrow \uparrow$ phases
in a narrow energy range accessible for the switching by
electric and magnetic fields.
It seems that
the available experimental data do not rule out this idea, although
at the present stage there is no direct
support to it either.
We hope that our work will stimulate theoretical and experimental activity in
this direction.

\ack
We are grateful to Alexei Belik for stimulating our interest to
the problem of magnetism and orbital ordering in BiMnO$_3$, numerous discussions,
and providing us unpublished structure parameters of BiMnO$_3$ at 4K.
We are also grateful to Tatsuya Shishidou for discussions
of unpublished results of his electronic structure calculations for BiMnO$_3$.
The work of IVS is partly supported by Grant-in-Aid for Scientific
Research in Priority Area ``Anomalous Quantum Materials''
and Grant-in-Aid for Scientific
Research (C) No. 20540337
from the
Ministry of Education, Culture, Sport, Science and Technology of
Japan.
The work of ZVP is partly supported
by Dynasty Foundation,
Grants of President of Russia MK-3227.2008.2, and scientific school
grant SS-1929.2008.2.


\section*{References}
\begin{thebibliography}{10}

\bibitem{Fiebig}
Fiebig M 2005
{\it J. Phys.} D: {\it Appl. Phys.} {\bf 38} R123

\bibitem{Khomskii}
Khomskii D I 2006
{\it J. Magn. Magn. Matter.} {\bf 306} 1

\bibitem{Eerenstein}
Eerenstein W, Mathur N D and Scott J F 2006
{\it Nature} {\bf 442} 759

\bibitem{Cheong}
Cheong S-W and Mostovoy M 2007
{\it Nature materials} {\bf 6} 13

\bibitem{belik_07}
Belik A A, Iikubo S, Yokosawa T, Kodama K, Igawa N,
Shamoto S, Azuma M, Takano M, Kimoto K, Matsui Y and Takayama-Muromachi E 2007
{\it J. Am. Cham. Soc.} {\bf 129} 971

\bibitem{Chiba}
Chiba H, Atou T, Syono Y 1997
{\it J. Solid State Chem.} {\bf 132} 139

\bibitem{HillRabe}
Hill N A and Rabe K M 1999
{\it Phys. Rev.} B {\bf 59} 8759

\bibitem{SeshadriHill}
Seshadri R and Hill N A 2001
{\it Chem. Mater.} {\bf 13} 2892

\bibitem{atou_99}
Atou T, Chiba H, Ohoyama K, Yamaguchi Y and Syono Y 1999
{\it J. Solid State Chem.} {\bf 145} 639

\bibitem{santos_02}
Moreira dos Santos A, Cheetham A K, Atou T, Syono Y,
Yamaguchi Y, Ohoyama K, Chiba H and Rao C N R 2002
{\it Phys. Rev.} B {\bf 66} 064425

\bibitem{montanari_05}
Montanari E, Calestani G, Migliori A, Dapiaggi M,
Bolzoni F, Cabassi R and Gilioli Ed 2005
{\it Chem. Mater.} {\bf 17} 6457

\bibitem{SantosSSC}
Moreira dos Santos A, Parashar S, Raju A R, Znao Y S,
Cheetham A K and Rao C N R 2002
{\it Solid State Commun.} {\bf 122} 49

\bibitem{Shishidou_04} Shishidou T, Mikamo N, Uratani Y, Ishii F and Oguchi T
2004 {\it J. Phys.: Condens. Matter} {\bf 16} S5677

\bibitem{Kimura}
Kimura T, Kawamoto S, Yamada I, Azuma M, Takano M and Tokura Y (2003)
{\it Phys. Rev.} B {\bf 67} 180401(R)

\bibitem{Sharan} Sharan A, Lettieri J, Jia Y, Tian W, Pan X,
Schlom D G, and Gopalan V 2004
{\it Phys. Rev.} B {\bf 69} 214109

\bibitem{montanari_07}
Montanari E, Calestani G, Righi L, Gilioli E, Bolzoni F,
Knight K S and Radaelli P G (2007)
{\it Phys. Rev.} B {\bf 75} 220101(R)

\bibitem{Yokosawa}
Yokosawa T, Belik A A, Asaka T, Kimoto K, Takayama-Muromachi E and Matsui Y 2008
{\it Phys. Rev.} B {\bf 77} 024111

\bibitem{EerensteinPML} Eerenstein W, Morrison F D, Sher F, Prieto J L,
Attfield J P, Scott J F,  Mathur N D 2007
{\it Phil. Mag. Lett.} {\bf 87} 249

\bibitem{belik_bisco3}
Belik A A, Iikubo S, Kodama K, Igawa N, Shamoto S,
Maie M, Nagai T, Matsui Y, Stefanovich S Yu, Lazoryak B I and Takayama-Muromachi E 2006
{\it J. Am. Chem. Soc.} {\bf 128} 706

\bibitem{Shishidou}
Shishidou T 2007
{\it private communication}

\bibitem{spaldin_07}
Baettig P, Seshadri R and Spaldin N A 2007
{\it J. Am. Chem. Soc.} {\bf 129} 9854

\bibitem{belik_pc} Belik 2007
{\it private communication}

\bibitem{Sergienko}
Sergienko I A, \c{S}en C and Dagotto E 2006
{\it Phys. Rev. Lett.} {\bf 97} 227204

\bibitem{Picozzi}
Picozzi S, Yamauchi K, Sanyal B, Sergienko I A and Dagotto E 2007
{\it Phys. Rev. Lett.} {\bf 99} 227201

\bibitem{Wang}
Wang C, Guo G-C and He L 2007
{\it Phys. Rev. Lett.} {\bf 99} 177202

\bibitem{LMTO1}
Andersen O K 1975 {\it Phys. Rev.} B {\bf 12} 3060

\bibitem{LMTO2}
Gunnarsson O, Jepsen O and Andersen O K 1983
{\it Phys. Rev.} B {\bf 27} 7144

\bibitem{LMTO3}
Andersen O K and Jepsen O 1984 {\it Phys. Rev. Lett.} {\bf 53} 2571

\bibitem{PRB06a}
Solovyev I V 2006
{\it Phys. Rev.} B {\bf 73} 155117

\bibitem{review2008}
Solovyev I V 2008
{\it J. Phys.: Condens. Matter}
(to be published)

\bibitem{PRB04}
Solovyev I V 2004
{\it Phys. Rev.} B {\bf 69} 134403

\bibitem{PRB07}
Solovyev I V, Pchelkina Z V and Anisimov V I 2007
{\it Phys. Rev.} B {\bf 75} 045110

\bibitem{BradlayCracknell}
Bradley C J and Cracknell A P 1972
\textit{The Mathematical Theory
of Symmetry in Solids} (Oxford: Clarendon Press)

\bibitem{MochizukiImada}
Mochizuki M and Imada M 2003
{\it Phys. Rev. Lett.} {\bf 91} 167203

\bibitem{PRB06b}
Solovyev I V 2006
{\it Phys. Rev.} B {\bf 74} 054412

\bibitem{Gunnarsson1989}
Gunnarsson O, Andersen O K, Jepsen O and Zaanen J 1989
{\it Phys. Rev.} B {\bf 39} 1708

\bibitem{Ferdi04}
Aryasetiawan F, Imada M, Georges A, Kotliar G, Biermann S
and Lichtenstein A I 2004
{\it Phys. Rev.} B {\bf 70} 195104

\bibitem{Liechtenstein}
Liechtenstein A I, Katsnelson M I, Antropov V P and Gubanov V A 1987
{\it J. Magn. Magn. Matter.} {\bf 67} 65

\bibitem{TRN}
Solovyev I V 2003
Magnetic Interactions in Transition-Metal Oxides
\textit{Recent Research Developments in Magnetism and Magnetic Materials}
(India: Transworld Research Network) vol 1 p 253

\bibitem{Springer}
Solovyev I V and Terakura K 2003
Orbital Degeneracy and Magnetism of Perovskite Manganese Oxides
({\it Electronic Structure and Magnetism of Complex Materials}),
ed D J Singh and D A Papaconstantopoulos (Berlin: Springer) p 253

\bibitem{PRL99}
Solovyev I V and Terakura K 1999
{\it Phys. Rev. Lett.} {\bf 82} 2959

\bibitem{Yang}
Yang C-H, Koo J, Song C, Koo T Y, Lee K-B and Jeong Y H 2006
{\it Phys. Rev.} B {\bf 73} 224112

\bibitem{Mazurenko}
Mazurenko V V, Skornyakov S L, Kozhevnikov A V, Mila F and Anisimov V I 2007
{\it Phys. Rev.} B {\bf 75} 224408

\bibitem{KugelKhomskii}
Kugel K I and Khomskii D I 1982
{\it Sov. Phys. Usp.} {\bf 25} 231

\bibitem{Oguchi}
Oguchi T, Ketarura K and Williams A R 1983
{\it Phys. Rev.} B {\bf 28} 6443

\bibitem{ZaanenSawatzky}
Zaanen J and Sawatzky G A 1987
{\it Can. J. Phys.} {\bf 65} 1262

\bibitem{deBrito}
de Brito P E and Shiba H 1998
{\it Phys. Rev.} B {\bf 57} 1539

\bibitem{PRB03}
Solovyev I V 2003
{\it Phys. Rev.} B {\bf 67} 014412

\bibitem{DMFT}
Georges A, Kotliar G, Krauth W and Rozenberg M J 1996
{\it Rev. Mod. Phys.} {\bf 68} 13

\bibitem{Nagaev}
Nagaev E L 1996
{\it Physics - Uspekhi} {\bf 39} 781

\bibitem{Dagotto}
Dagotto E 2005
{\it New J. Phys.} {\bf 7} 67

\bibitem{Kodama}
Kodama K, Iikubo S, Shamoto S-i, Belik A A and Takayama-Muromachi E 2007
{\it J. Phys. Soc. Jpn.} {\bf 76} 124605

\bibitem{Konishi}
Konishi Y, Fang Z, Izumi M, Manako T, Kasai M, Kuwahara H,
Kawasaki M, Terakura K and Tokura Y 1999
{\it J. Phys. Soc. Jpn.} {\bf 68} 3790

\bibitem{Fang}
Fang Z, Solovyev I V and Terakura K 2000
{\it Phys. Rev. Lett.} {\bf 84} 3169

\bibitem{Ohshima}
Ohshima E, Saya Y, Nantoh M and Kawai M 2000
{\it Solid State Commun.} {\bf 116} 73

\bibitem{Uehara}
Uehara M, Mori S, Chen C H and Cheong S-W 1999,
{\it Nature} {\bf 399} 560

\end{thebibliography}
\end{document}